\documentclass[12pt]{article}
\usepackage[utf8]{inputenc}
\usepackage{graphicx}
\usepackage{epsfig}
\usepackage{amsmath}
\usepackage{amsthm,amssymb}
\usepackage[round]{natbib}
\usepackage{color}
\usepackage{xcolor,colortbl}
\usepackage{rotating,url}
\usepackage[normalem]{ulem}
\usepackage{units}
\usepackage{multirow}

\newcommand{\n}[1] {\mbox{\boldmath{$#1$}}}

\newcommand{\btheta}{\boldsymbol{\theta}}

\newcommand{\CM}{f}
\newcommand{\EM}{e}

\newtheorem{theorem}{Theorem}
\newcommand{\be}{\begin{eqnarray}}
\newcommand{\ee}{\end{eqnarray}}
\newcommand{\beq}[1]{\begin{equation}\label{#1}}
\newcommand{\eeq}{\end{equation}}
\newcommand{\ba}{\begin{eqnarray*}}
\newcommand{\ea}{\end{eqnarray*}}

\DeclareMathOperator{\PIPS}{PIPS}

\newcommand{\dirac}{\text{Dir}}

\definecolor{Gray}{gray}{0.85}
\definecolor{LightCyan}{rgb}{0.88,1,1}

\newcolumntype{a}{>{\columncolor{Gray}}c}
\newcolumntype{b}{>{\columncolor{white}}c}

\newcommand{\blind}{0}
\title{Screening the Discrepancy Function of a Computer Model}
\author{Pierre Barbillon\thanks{Université Paris-Saclay, AgroParisTech, INRAE, UMR MIA Paris-Saclay, 91120, Palaiseau, France, \texttt{pierre.barbillon@agroparistech.fr}},
Anabel Forte\thanks{Universitat de Valencia, \texttt{anabel.forte@uv.es}.}  
\ and Rui Paulo
\thanks{CEMAPRE/REM and Department of Mathematics, Lisbon School of Economics and Management, Universidade de Lisboa, \texttt{rui@iseg.ulisboa.pt}}.}
\date{}

\begin{document}



\if0\blind
{
  \title{\bf Screening the Discrepancy Function of a Computer Model}
  \author{Pierre Barbillon\hspace{.2cm}\\
    Université Paris-Saclay, AgroParisTech, INRAE, UMR MIA Paris-Saclay,\\ 91120, Palaiseau, France\\
    and \\
    Anabel Forte \\
    Department of Statistics and Operations Research, Universitat de Valencia\\
    and
    \\
    Rui Paulo \\
    CEMAPRE/REM and Department of Mathematics, \\ Lisbon School of Economics and Management, Universidade de Lisboa
    }
  \maketitle
} \fi

\if1\blind
{
  \bigskip
  \bigskip
  \bigskip
  \begin{center}
    {\LARGE\bf Screening the Discrepancy Function of a Computer Model}
\end{center}
  \medskip
} \fi

\vspace{-1cm}
\begin{abstract}
Traditionally, screening refers to the problem of detecting influential (active) inputs in the computer model. We develop methodology that applies to screening, but the main focus is on detecting active inputs not in the computer model itself but rather on the discrepancy function that is introduced to account for model inadequacy when linking the computer model with field observations. We contend this is an important problem as it informs the modeler which are the inputs that are potentially being mishandled in the model, but also along which directions it may be less recommendable to use the model for prediction. The methodology is Bayesian and is inspired by the continuous spike and slab prior popularized by the literature on Bayesian variable selection. In our approach, and in contrast with previous proposals, a single MCMC sample from the full model allows us to compute the posterior probabilities of all the competing models, resulting in a methodology that is computationally very fast. The approach hinges on the ability to obtain posterior inclusion probabilities of the inputs, which are easy to interpret quantities, as the basis for selecting active inputs. For that reason, we name the methodology PIPS --- posterior inclusion probability screening.
\end{abstract}

\noindent%
\vfill

\newpage

\section{Introduction}\label{intro}

Understanding the complexities of the world and its intricate processes has always posed a challenge for humankind. Over centuries, scientific research has allowed us to approximate these processes by creating mathematical or statistical models that combine parameters and variables. In this work, we specifically focus on deterministic mathematical models, commonly known as computer models, which attempt to simulate these processes of interest. However, it's important to note that such models only represent our limited knowledge of the actual phenomenon. Therefore, a crucial aspect is validation, a process that involves comparing the model with field data, which consists of potentially noisy measurements of the real phenomenon \citep{BayarriTech2007}. 

In this context, there are multiple sources of uncertainty  \citep{KO2001}. To name but a few: noise on the field measurements, unknown parameters in the computer model, 
and the mismatch between the computer model and reality. This last source of uncertainty, also known as the discrepancy or the bias of the computer model, is usually incorporated within the statistical framework that links the field data with the computer model. Moreover, it is commonly modeled as a Gaussian Stochastic Process (GaSP) \citep{KO2001,higdon2004combining,BayarriTech2007}.

The objective of this work is to explore the discrepancy function as a tool to help modelers understand the shortcomings of the computer model and identify areas that require improvement. In particular, our GaSP model for the discrepancy depends on, at least, the same input variables as the ones included in the computer model. Then,  by screening the discrepancy function to determine the influential inputs, which we call \emph{active}, we hopefully identify the potential flaws of the computer model. This approach aligns with the work of \cite{JosephYan} but we use a single step approach --- in the spirit of a fully Bayesian analysis as in \cite{BayarriTech2007}. Our methodology, named PIPS (posterior inclusion probability screening), relies on obtaining posterior inclusion probabilities of each of the inputs. These probabilities are straightforward to interpret and serve as the basis for labeling a variable as active.

To implement this procedure, we resort to the idea of spike and slab prior distributions \citep{MB1988}, which are commonly used for selecting variables in
linear regression models \citep{bai2020spike}. We use essentially the same idea to model the range parameters involved in the correlation function of the GaSP for the discrepancy function, taking advantage of the reparametrization proposed in \citet{linketal2006}, which puts the range parameters in the compact interval $[0,1]$ with the value $1$ formally corresponding to an inert input, i.e., not active.

After setting prior distributions, a method for variable selection is needed. To the best of our knowledge, two main approaches have been used for performing variable selection within the GaSP: 
\begin{itemize}
\item the reference distribution variable selection (RDVS) method,  where a fictitious variable is artificially added to the GaSP and an MCMC algorithm is performed to sample from the posterior of the range parameters. The posterior distribution of the parameters associated with the input variables are then compared with that associated with the fictitious variable in order to determine which are active and which are not \citep{linketal2006};

\item high dimensional MCMC algorithms \citep{savetal2011,chen2010bayesian}, where the exploration step goes through the different models, each with potentially different numbers of active inputs. This type of methods provides a posterior probability of each input variable being active, calculated as the proportion of times a model which includes this input was visited in the MCMC.  This approach has the advantage of being formally easier to justify. However, it is  computationally demanding and the MCMC schemes that visit models with varying dimensions are often hard to tune.
\end{itemize}

Instead, in this work we propose an approach relying on a smooth version of the spike and slab prior distribution. We only need to run a single MCMC algorithm under the model where the slab  prior is chosen for all the input variables. Then, a post-processing of the resulting sample allows us to obtain the posterior probability of each input variable  being active in the discrepancy function. 


\paragraph{Outline of the paper}
In Section \ref{bkgrd}, we provide the background and establish the statistical model as well as the notation considered in the rest of the paper.  
Then, in Section \ref{met}, we describe in detail our proposal, including a critical discussion of the methodological choices that frame it. Section \ref{simul} is devoted to synthetic
examples as a means to illustrate the potential benefits brought by PIPS in the realm of uncertainty quantification. Section \ref{sec:realapp}
shows a real example where we screen the discrepancy function of a computer model that models the electrical production
of a photovoltaic plant. Finally, Section \ref{conclusion} concludes the paper with some general comments and directions for future research. Throughout the paper, we relegate to the Supplementary Material aspects of the discussion that are perceived as not essential to the development of the methodology.


\section{Background, notation and assumptions}\label{bkgrd}

\subsection{Calibration and model inadequacy}\label{calibration}
Computer models typically have two kinds of inputs, namely, input variables and model parameters. Input variables are often controllable inputs which can be set at chosen values in field experiments designed to observe the real process which the computer model aims at reproducing. They may also be environmental variables which are observed as part of the field experiments. On the other hand, model parameters are usually unknown in the context of physical experiments; they are typically estimated using the information contained in the physical experimental data, a task that is usually called calibration. 

\cite{KO2001} described the various sources of uncertainty that are present in the process of linking experimental observations and computer models, often for the purpose of calibration. One of these sources of uncertainty is referred to by these authors as model inadequacy and results from the fact that no model is a perfect representation of the real process it aims at reproducing. They propose that this uncertainty should be captured by the so-called discrepancy or bias function; \cite{craig1996,craig1997pressure,craigetal2001} advocate a similar approach. Since then, this has become standard practice, although various cautionary remarks have been made over the years as to how difficult it is to account for this extra source of uncertainty, cf. e.g. \cite{Brynjarsdottir_2014}. 

Specifically, we denote the generic vector of $p$ input variables by $\n x=(x_1\ldots,x_{p})^\top$ and the generic vector of $k$ model parameters by $\n \theta=(\theta_1,\ldots,\theta_{k})^\top$. Then, $\n x_1,\ldots,\n x_n$ denote the $n$ values of $\n x$ at which the field experiments are conducted (if controlled), or the $n$ values of $\n x$ that will be observed as part of each of the $n$ experiments (if corresponding to environmental variables); that is, $\n x_i=(x_{1i},\ldots,x_{pi})^\top$ denotes the values of the input variables for the $i$th field experiment. Following \cite{KO2001}, we model the field data as
\begin{equation}\label{KOH}
y(\n x_i) = \CM(\n x_i,\n \theta) + \delta(\n x_i)+\varepsilon_i
\end{equation}
where, for $i=1,\ldots,n$,  $\varepsilon_i$ are independent $\mathcal{N}(0,\sigma_0^2)$ random variables which represent measurement error, $\CM(\cdot,\cdot)$ denotes the computer model and $\n \theta$ represents the true but unknown value of the vector of model parameters. 
The discrepancy function, $\delta(\cdot)$, is modeled \textit{a priori} as a GaSP:
\begin{equation}\label{priordelta}
\delta(\cdot)\mid \sigma^2, \n \psi \sim \text{GaSP}(0, \sigma^2 c(\cdot,\cdot\mid \n \psi))
\end{equation}
where $c(\cdot, \cdot\mid \n \psi)$ is chosen as a separable correlation kernel which, for any two configurations $\n x_i$ and $\n x_j$, is given by
\begin{equation}\label{sep}
c(\n x_i,\n x_j \mid \n \psi)=\prod_{\ell=1}^{p} c(x_{\ell i},x_{\ell j}\mid \psi_\ell)
\end{equation}
and $\psi_\ell\in (0,+\infty)$ denotes a range parameter. One of the most common choices for $c$ is the power exponential kernel
\begin{equation}\label{pwex}
c(x_{\ell i},x_{\ell j}\mid \psi_\ell)=\exp(-|x_{\ell i}-x_{\ell j}|^a/\psi_\ell)
\end{equation}
with $0< a\leq 2$ fixed. We will use this kernel throughout the paper but reparametrized, c.f. Section~\ref{screen}. Mainly for computational reasons, that is, to avoid nearly-singular correlation matrices, $a$ will be fixed at $a=1.9$. Notice that we impose a zero mean for this Gaussian process.  This is justified by the belief that the computer model accounts for the main trend in the field data.

In this setting, the sampling distribution of the field data $\n y^\top=(y_1,\ldots,y_n)$ is such that, with $\n \CM(\n \theta)= (\CM(\n x_i,\n \theta),\ i=1,\ldots,n)^\top$ ,
\begin{equation}\label{lik2}
\n y  \mid \n \psi, \sigma^2, \sigma_0^2,\n \theta, \n \CM(\n \theta) \sim \mathcal{N}_n(\n \CM(\n \theta), \sigma^2 \n R + \sigma_0^2\ \n I_n)
\end{equation}
where $\n R$ is a $n\times n$ matrix with entries $\n R=[c(\n x_i,\n x_j\mid\n \psi)]_{i,j=1,\ldots,n}$ and $\n I_n$ denotes the order-$n$ identity matrix.

In the situation where $\CM(\cdot,\cdot)$ is computationally  fast, the statistical model \eqref{lik2}, paired with prior distributions on the unknown parameters $\n \theta$, $\sigma^2$, $\n \psi$ and $\sigma_0^2$, allows us to obtain the posterior distribution of the vector of model parameters $\n \theta$. Additionally, it also allows us to predict the real process at particular configurations of the input variables. Such predictions incorporate model discrepancy and are, for that reason, called bias-corrected predictions in \cite{BayarriTech2007}. 

The focus of this paper is on how to use this same Bayesian statistical set-up to ascertain which of the input variables $x_1,\ldots,x_{p}$ are active in the discrepancy function, in a sense that will be made precise in the sequel. By doing so, we will determine which variables need to be taken into account in the bias term and, hence, are not handled correctly by the computer model. This is also a valuable insight for the potential users of the model, as extrapolating in the direction of an active variable can be hazardous.

In the next subsection, we present methods for screening the discrepancy function --- i.e., for determining which variables are active ---  based on the existing literature. Although this literature is concerned with screening a GaSP emulator \citep{linketal2006} or with variable selection for GaSP regression \citep{savetal2011}, screening the discrepancy function when the prior \eqref{priordelta} is used and $f(\cdot,\cdot)$ is computationally fast is actually very similar to the problem at hand. In fact, the statistical model on which screening a computer model is based is identical to \eqref{lik2} but with $\n \CM(\n\theta)\equiv \n0$ and $\delta$ being the prior on the computer model. In that case, $\n y$ is the so-called model data, obtained by running the computer model at a set of carefully designed configurations for its inputs. Additionally, $\sigma^2$  corresponds to  the variance of the GaSP prior assumed for the computer model and $\sigma_0^2$ is the variance of the nugget --- see \cite{GL2012} for reasons why a nugget may be used when emulating deterministic models. 

Of course, $f(\cdot,\cdot)$ is very often computationally demanding. We will address this situation in Section~\ref{ssec:slowcc}.

\subsection{Screening the discrepancy function}\label{screen}
In the context of the model given by \eqref{lik2},  the statement that one of the  $p$ input variables, say $x_\ell$, does not affect the discrepancy function simply means that, as $x_\ell$ varies, the correlation matrix $\n R$ does not change. In that case, we label $x_\ell$ as \emph{inert}; otherwise, we label it as \emph{active}.
Given the separable structure that we have assumed for the correlation function, as detailed in \eqref{sep}, this is expressed in terms of a single parameter, namely, $\psi_\ell$, the range parameter. Typically, as $\psi_\ell\rightarrow +\infty$, and for any $i, j=1, \ldots, n$, $i\neq j$, we have $c(x_{\ell i}, x_{\ell j}\mid \psi_\ell)\rightarrow 1$, and hence $x_\ell$ does not contribute to $\n R$; it is inert. This can be verified in the particular case of the power exponential correlation function in \eqref{pwex}, which is very popular in the area of computer models and that we assume throughout this paper but with a convenient reparametrization. That is, instead of the range parameter, we use the parametrization introduced by \cite{linketal2006} specifically to address the problem of screening of computer models: $\rho_\ell=\exp(-(1/2)^a/\psi_\ell)$, that is, we use:
\begin{equation}\label{rho}
c(x_{\ell i}, x_{\ell j}\mid \rho_\ell)=\rho_\ell^{(2 |x_{\ell i}-x_{\ell j}|)^{a}}\ .
\end{equation}

There are two advantages to this parametrization: first, the $\ell$-th input is inert if $\rho_\ell=1$; second, the parameter space for $\rho_\ell$ is the unit interval, which simplifies thinking about priors for this parameter, particularly when the inputs are also transformed to vary in $[0,1]^p$, as will be the case throughout this paper. As for other correlation kernels, see Section~\ref{otherck}.

It is natural to pose the problem of screening as a variable selection problem and proceed by indexing the models that result from selecting all possible subsets of active input variables using the vector $\n \gamma=(\gamma_1,\ldots,\gamma_p)^\top\in\{0,1\}^p$. The screening exercise can now be cast as assessing the evidence in favor of each of the competing models. Under model $\mathcal{M}_{\boldsymbol \gamma}$, the sampling distribution of the observable $\n y$ is as in \eqref{lik2} except for the fact that $\n R$ is replaced by $\n R_{\boldsymbol \gamma}$, the entries of which are computed using the set of $\rho_\ell$ such that $\gamma_\ell=1$, $\ell=1,\ldots,p$. Equivalently, under  $\mathcal{M}_{\boldsymbol \gamma}$, $\rho_\ell=1$ if $\gamma_\ell=0$.
This language is very similar to the one used in Bayesian variable selection in the context of the linear regression model \cite[cf, e.g.,][]{CG2004}. 

A natural way to quantify model uncertainty in this context is through the posterior distribution on the model space,
\begin{equation}\label{postprob}
\pi(\boldsymbol \gamma\mid \n y) \propto m(\n y\mid \n \gamma)\ \pi(\n \gamma)
\end{equation}
where $\pi(\n \gamma)=\mathbb{P}(\mathcal{M}_{\boldsymbol \gamma})$ and $\pi(\boldsymbol \gamma\mid \n y) = \mathbb{P}(\mathcal{M}_{\boldsymbol \gamma}\mid \n y)$ denote, respectively, the prior and posterior probabilities of model $\mathcal{M}_{\boldsymbol \gamma}$ and $m(\n y\mid \n \gamma)$ represents the prior predictive distribution of the observable $\n y$ under that model, that is,
\begin{equation}\label{marg2}
m(\n y\mid \n \gamma) =\int \mathcal{N}(\n y\mid \n \CM(\n\theta), \sigma^2\ \n R_{\boldsymbol \gamma}+ \sigma^2_0\ \n I_n)\ \pi(\sigma^2, \sigma_0^2,\n \rho\mid \n \gamma)\ \pi(\n\theta)\ d\sigma^2\ d\sigma_0^2\ d \n \rho\ d\n\theta\ ,
\end{equation} 
where we denote by $\mathcal{N}(\cdot\mid \n \mu,\n \Sigma)$ the density of a multivariate normal distribution with mean vector $\n \mu$ and covariance matrix $\n \Sigma$. We also refer to $m(\n y\mid \n \gamma)$ as the marginal likelihood under model $\n \gamma$.

Besides the integration present in \eqref{marg2}, another difficulty in this approach is the specification of the prior on the model specific parameters, $\pi(\sigma^2, \sigma_0^2, \n \rho\mid \n\gamma)$. The prior on the calibration parameters $\n \theta$, however, is typically specified using expert information.

\cite{savetal2011} extends the work initiated by \cite{linketal2006} for the screening of computer models to more general data structures and models. They propose to use $\pi(\sigma^2, \sigma_0^2, \n \rho\mid \n\gamma)=\pi(\sigma^2)\ \pi(\sigma_0^2)\ \pi(\n \rho\mid \n\gamma)$ where
\begin{align} \label{rhosav}
\pi(\n\rho\mid \n \gamma) = \prod_{\ell=1}^p \left[\gamma_\ell\ \mathcal{U}(\rho_\ell) + (1-\gamma_\ell)\ \dirac_1(\rho_\ell)\right]
\end{align}
with $\dirac_1(\cdot)$ representing the Dirac delta at 1 and $\mathcal{U}(\cdot)$ denoting the uniform density in the $(0,1)$ interval. This is inspired by the (discrete) spike and slab prior of Bayesian variable selection \citep{MB1988}: if a variable $x_\ell$ is present in the model, the prior for $\rho_\ell$ is the `slab', a $\mathcal{U}(0,1)$ here; otherwise it's a `spike', a point mass at 1. [Note that placing a uniform prior on $\rho_\ell$ translates to an Inverse-Gamma distribution on the range parameter $\psi_\ell$.] The prior \eqref{rhosav} is paired with 
$$\pi(\n \gamma) = \prod_{\ell=1}^p {\tau_\ell}^{ \gamma_\ell} (1-\tau_\ell)^{ 1-\gamma_\ell}\ ,$$
where $\tau_\ell$ is a fixed number representing the prior probability that $x_\ell$ is active. They additionally propose fairly sophisticated MCMC schemes to sample from the posterior distribution of $(\n \rho, \sigma^2, \sigma_0^2, \n\gamma)$. The selection of variables is made by inspecting the posterior on $(\n \rho,\n \gamma)$. 

The approach in \cite{linketal2006} is different, but related.
They consider a prior on $\n \rho$ that can be obtained from the set-up described above by setting $\tau_\ell=\tau$ for all $1\le \ell\le p$ and then integrating out $\n \gamma$ from $\pi(\n \rho,\n\gamma)= \pi(\n\rho\mid \n \gamma)\ \pi(\n \gamma)$, resulting in 
\begin{equation}\label{rholink}
\pi(\n \rho)  = \prod_{\ell=1}^p \left[\tau\ \mathcal{U}(\rho_\ell) + (1-\tau)\dirac_1(\rho_\ell)\right]\ .
\end{equation}
Since the model indicator $\n \gamma$ is integrated out, variable selection can only be based on the posterior distribution of $\n \rho$, and one needs to find a criterion to determine whether $x_\ell$ is active. The process leading to that decision is quite involved: for a large number of times (say, $T=100$), add a fictitious input $x_{\text{new}}$ to the correlation kernel (along with $\rho_{\text new}$) and to the design set, obtain the posterior distribution of $(\n \rho,\rho_{\text{new}})$, record the posterior median of $\rho_{\text{new}}$. In the end, deem an input $x_\ell$ as inert if the posterior median of $\rho_\ell$ 
exceeds a fixed lower percentile (say, the 10\%) of the distribution of the posterior median of $\rho_{\text{new}}$. The need for this comparison justifies the acronym of their approach: RDVS, or reference distribution variable selection.  Note that the price to pay for avoiding a sophisticated MCMC algorithm to sample from $\n\rho, \n \gamma,\sigma^2,\sigma_0^2 \mid \n y$ is to repeat a standard (i.e., fixed dimension) MCMC algorithm $T$ times.

The methods proposed in \citet{savetal2011} and \citet{linketal2006} are computationally demanding since they resort, respectively, either to a high-dimensional MCMC algorithm or to many repetitions of an MCMC algorithm.

Alternative approaches to assessing the influence of input variables in the context of a GaSP regression include those of \cite{paananen2019variable} and \citet[Section 3.5]{lee2017prediction}. In common, these references have the fact that they are both concerned with ranking the input variables according to their impact on the variability of the output and not with classifying them as active/inert. Also,  both references rely on plug-in estimates of the range parameters of the GaSP, neglecting parameter uncertainty. Although these papers propose an interesting link between the values of the range parameters and the importance of the associated input variables, it is difficult to include the resulting methods in our fully-Bayesian variable selection procedure.

\section{The PIPS methodology}\label{met}

In the previous section, we have laid out the problem of screening 
the discrepancy function 
and  have cast the screening problem as a model selection exercise.
Here, we detail the methodology that we propose to solve this problem. 
We first present the class of competing
models and then we expose our methodology to compute their posterior probabilities. 
Next, we describe how to extend the methodology in order to address the problem when the computer model is slow. Lastly, we discuss and further justify methodological choices that frame our proposal.

\subsection{The class of competing models and prior distributions}\label{subsec:prior}

Let $\n \eta= (\sigma^2,\sigma_0^2, \n \theta)^\top$ be the vector containing 
the variance of the GaSP prior of $\delta$, 
the variance of the measurement error, and the vector of calibration parameters.
The vector $\n \gamma$, as defined in the previous section, indexes all the possible selections of active input variables.
The issues that we have to address are the specification of the prior on the vectors $\n \eta$ and $\n \rho$ under each model $\mathcal{M}_{\boldsymbol \gamma}$, and the computational problem of obtaining the marginals in \eqref{marg2}. 

The prior in \eqref{rhosav} is reminiscent of the discrete spike and slab prior of variable selection in the linear regression context; see \cite{MB1988}. The continuous spike and slab, originally proposed by \cite{GM93}, differs from its discrete version in that the `spike' is no longer a Dirac delta but rather a continuous distribution highly concentrated around the value of the parameter that corresponds to removing the variable from the model, zero in the case of linear regression. 

We modify \eqref{rhosav} in the spirit of the continuous spike and slab. Hence, it is no longer the case that each of the competing models has a different number of parameters; rather, there is only one statistical model and what changes is the prior. We can still use the vector $\n \gamma$ to index the different selections of active inputs, but now $\gamma_\ell=0$ no longer means $\rho_\ell=1$ but rather that $\rho_\ell$ is highly concentrated around 1, which is interpreted as the impact of the input being of no practical significance, i.e., that it  is inert from a practical point of view. The quantities in \eqref{postprob}  are still well defined, and form the basis for posterior inference. For that reason, we continue to identify different values of $\n \gamma$ with different competing models. 

We introduce this modification for computational reasons, as it will allow us to compute all $2^p$ posterior probabilities using only an MCMC sample from the posterior of unknowns under the full model, the one corresponding to $\n \gamma=\n 1$, plus some minimal additional post processing --- this is detailed in Section~\ref{postcomp}.

Specifically, we propose to use $\pi(\n\rho,\n \eta, \n \gamma)=\pi(\n \eta)\ \pi(\n\rho\mid  \n \gamma)\ \pi(\n \gamma)$ where
\begin{equation} \label{rhoours}
\pi(\n\rho\mid \n \gamma) = \prod_{\ell=1}^p \left[\gamma_\ell\ \mathcal{U}(\rho_\ell) + (1-\gamma_\ell)\  \mathcal{B}(\rho_\ell\mid \alpha,1)\right]
\end{equation}
where $\mathcal{B}(\cdot\mid \alpha, \beta)$ represents the Beta density 
with positive shape parameters $\alpha$ and $\beta$. In \eqref{rhoours}, 
$\alpha$ is a large value, so that if an input variable is active then its corresponding $\rho$ 
follows \textit{a priori} a uniform distribution; otherwise, instead 
of being set at 1 like in the discrete spike and slab, it follows a 
distribution highly concentrated around 1. 
Since the inputs are transformed to vary in $[0,1]^p$, it suffices to consider the situation where the parameters of the `slab' are all the same. The choice of $\alpha$ is critical because it essentially determines how large a $\rho_\ell$ needs to be in order for the corresponding input variable to be considered inert from a practical point of view. In Section~\ref{choicealpha} we construct a heuristic argument to support the choice $\alpha=5000$, which is used in all the calculations of this paper.


In the following theorem we establish that, as
$\alpha\rightarrow +\infty$, the resulting inference approaches what is obtained via the discrete spike and slab prior, so that this construction
can be viewed as a relaxation technique to facilitate the sampling that we are 
proposing. Precisely, 

\begin{theorem} Fix $\ell\in \{1,\ldots,p\}$. 
Consider inference under the discrete spike and slab and assume that $\gamma_\ell=0$. Denote by $m^D(\n y\mid \n \gamma)$ the corresponding marginal likelihood. If we replace the prior on $\rho_\ell$ (a point mass at 1) by a $\mathcal{B}(\rho_\ell\mid \alpha,1)$, and denote by $m_\alpha(\n y\mid \gamma)$ the corresponding marginal likelihood, then, 
under some minor conditions on the prior distributions,
it holds that
\begin{equation}
 \lim_{\alpha\rightarrow\infty}m_\alpha(\n y\mid \n \gamma)=m^D(\n y\mid \n \gamma)\ .
\end{equation}
\end{theorem}


The proof and the minor conditions are provided in the Supplementary Material, Section~1.

The prior on $\n \gamma$ is not the subject of this paper. 
It is typically used to control for multiplicity \citep{ScottBerger09}, an issue that, 
to the best of our knowledge, has not been addressed in the context of 
screening. Once the marginal likelihoods have been computed, 
it is straightforward to include any prior on $\n\gamma$ via \eqref{postprob}. 
In our numerical examples, the constant prior will be the implemented choice, i.e., the prior under which all $2^p$ models are assigned the same probability \textit{a priori}.

The prior on $\n \eta$ is broken down as $\pi(\n \eta)=\pi(\n \theta)\ \pi(\sigma^2)\ \pi(\sigma_0^2)$. Expert knowledge is usually utilized to specify $\pi(\n \theta)$.  Inverse gamma distributions are chosen  for the priors on the variances; see sections \ref{simul} and \ref{sec:realapp} for examples on how to specify the parameters of these priors. 
Then, we need an efficient way to compute the marginal likelihood for all possible choices of $\n \gamma$ under the prior construction that we have just outlined, which we henceforth denote by  $m(\n y\mid \n \gamma)$ . That is the subject of the next section.

\subsection{Posterior inclusion probability computation}\label{postcomp}

The ability to compute the prior predictive distribution of the observable under each of the competing models is one of the main computational obstacles to implementing model selection strategies. \cite{HC2001} offers a comparative review of most of the available tools to accomplish this task. Many are based on some post-processing of a sample from the posterior distribution of the model-specific parameters, and this means that one would need an MCMC sample from each of the competing models, which is not feasible even for relatively small $p$. Our problem has a very specific structure that can be exploited.

First, notice that the posterior distribution over the model space in \eqref{postprob} can be written as a function of the Bayes factors of
each competing model to $\mathcal{M}_{\boldsymbol 1}$: 
\begin{equation}\label{Bgamma1}
B_{\boldsymbol \gamma} = \frac{m(\n y\mid \n \gamma)}{m(\n y\mid \n \gamma=\n 1)}
\end{equation}
which is a ratio of normalizing constants.
Since the support of the posterior is the same under model $\mathcal{M}_{\boldsymbol \gamma}$ and under model $\mathcal{M}_{\boldsymbol 1}$, 
we can use a version of importance sampling 
\citep[c.f., e.g.,][]{CS1997} to write this ratio of normalizing constants as an expectation of a ratio with respect to the posterior under $\mathcal{M}_{\boldsymbol 1}$, the full model. Specifically, 
\begin{equation}\label{Pips}
B_{\boldsymbol \gamma} = E_{\boldsymbol 1}\left[\frac{g(\n y\mid \n \rho, \n \eta, \n \gamma)\ \pi(\n \rho, \n \eta \mid \n\gamma)}{g(\n y\mid \n \rho, \n \eta, \n \gamma=\n 1)\ \pi(\n \rho, \n \eta \mid \n\gamma=\n 1)}\right]\ ,
\end{equation}
where $g(\n y\mid \n \rho, \n \eta, \n \gamma)$ denotes the sampling density of $\n y$ that results from \eqref{lik2} under model $\mathcal{M}_{\boldsymbol \gamma}$. To clarify: in \eqref{Pips}, for each fixed $\n \gamma$, the expectation is taken with respect to the distribution of  $(\n \rho, \n \eta)$ under the model where all inputs are active, i.e., $\n \gamma=\n 1$. In the numerator of the fraction, the prior density and the sampling density are computed under the model corresponding to the fixed $\n \gamma$, whereas in the denominator those functions are computed under model $\n \gamma=\n 1$. This allows us to estimate all the $2^{p}-1$ ratios \eqref{Bgamma1} using a single MCMC sample obtained from the posterior distribution of the unknowns under the full model. That is, if $\{\n \rho^{(r)},\n \eta^{(r)},\ r=1,\ldots,M\}$ is a sample from the posterior distribution of the unknowns when $\n \gamma=\n 1$, then
\begin{align}\label{ISE}
\begin{split}
B_{\boldsymbol \gamma}& \approx \frac{1}{M} \sum_{r=1}^M\
\frac{g(\n y\mid \n \rho^{(r)}, \n \eta^{(r)}, \n \gamma)\ \pi(\n \rho^{(r)}, \n \eta^{(r)} \mid \n\gamma)}{g(\n y\mid \n \rho^{(r)}, \n \eta^{(r)}, \n \gamma=\n 1)\ \pi(\n \rho^{(r)}, \n \eta^{(r)} \mid \n\gamma=\n 1)}\\
 &=\frac{1}{M} \sum_{r=1}^M \ \pi(\n \rho^{(r)} \mid \n\gamma)
\end{split}
\end{align}
which is a consequence of \eqref{lik2} being the same regardless of $\n \gamma$, of the prior independence between $\n \rho$ and $\n \eta$ given $\n \gamma$, and of the fact that under the full model the prior on $\n \rho$ is a product of uniform distributions on the unit interval.

We relegate to the Supplementary Material, Section~2 the details on how to obtain \eqref{Pips} and additional properties of the resulting estimator. In particular, we establish that \eqref{ISE} corresponds to a finite variance importance sampling estimator.

Once all $2^p-1$ Bayes factors are computed, it is straightforward 
to pair them with a prior on $\n \gamma$, e.g. the prior that assigns the same probability to all competing models, and to obtain the posterior probabilities of each of the competing models, $\pi(\n y\mid \n \gamma)$.
We can then compute
the so-called posterior (marginal) inclusion probability of each input $x_\ell$
\begin{equation}\label{incprob}
\PIPS(x_\ell) = \mathbb{P}(\{\mathcal{M}_{\boldsymbol \gamma}:\gamma_\ell=1\}\mid \n y)=\sum_{\boldsymbol \gamma:\ \gamma_\ell=1}\ \pi(\boldsymbol \gamma\mid \n y) 
\end{equation}
which is an easy to interpret measure of the importance of $x_\ell$ in explaining the response. In essence, our approach to screening reduces to the computation of these quantities.

We can additionally obtain other interesting summaries, e.g.\ the posterior inclusion probability of a pair of variables:
\begin{equation}\label{incprobpair}
\PIPS(x_\ell\vee x_ j) = \PIPS(x_\ell)+\PIPS(x_j)-\sum_{\boldsymbol \gamma:\, \gamma_\ell=1, \gamma_j=1}\ \pi(\boldsymbol \gamma\mid \n y)\ .
\end{equation}
 In Section~\ref{sec:realapp}, we illustrate a possible use for the posterior inclusion probability of a pair of variables.
 
\subsection{Slow computer code}
\label{ssec:slowcc}

When the computer model $f(\cdot,\cdot)$ is computationally too expensive to render the MCMC algorithms unfeasible, a classical technique is to bypass the computational effort by replacing the computer model with a surrogate which approximates its output but is fast to run.
Thus, the surrogate can be plugged in the likelihood derived from the model given in 
 \eqref{lik2} instead of the actual computer model, represented by $\n \CM(\n \theta)$. However, the main limitation of this approach lies in the fact that the bias not only accounts for model discrepancy but also for the inaccuracies that result from the use 
of the  surrogate as an approximation to the computer model.

Some surrogates are what we call emulators with, perhaps, the most utilized being  the GaSP emulator \citep{sacks1989design,currin1991bayesian}. This is an interesting choice since, in addition to a
fast approximation, the emulator provides the practitioner with a measure of the additional source of 
uncertainty stemming from the approximation.

Specifically, the idea is to consider a limited number of runs of the computer model for a set of values $\n D=\{(\n x^\star_i,\boldsymbol \theta_i), i=1,\ldots, n_D\}$ which consists in a series of designed combinations of the values of the input variables and model
parameters. We then denote by $\CM(\n D)$ the corresponding outputs of the computer model for the combinations in $\n D$. Note that the configurations chosen for the input variables, $\{\n x_i^\star, i=1,\ldots, n_D\}$, may differ from the ones corresponding to the field data, $\{\n x_i, i= 1,\ldots, n\}$.

In this approach, the computer model is then modeled as a GaSP using the principle of modularization \citep{liu2009modularization}. This principle is based on the idea of estimating the GaSP parameters using only  $\CM(\n D)$, effectively discarding the field data, $\n y$, in this process.
The emulator then consists of the process conditioned 
on $\CM(\n D)$ with plugged-in estimates of the parameters. The 
conditional mean of this GaSP, denoted by $\EM(\cdot)$,  provides an approximation 
of the computer model while its conditional covariance, denoted by $k(\cdot,\cdot)$, 
provides a measure on the uncertainty stemming from this approximation.  
This source of uncertainty can be incorporated within the likelihood/statistical model:
\begin{equation}\label{lik2emulator}
\n y  \mid \n \psi, \sigma^2, \sigma_0^2,\n \theta, \CM(\n D) \sim N_n(\n \EM(\boldsymbol \theta), \sigma_f^2\n K + \sigma^2 \n R + \sigma_0^2\ \n I_n)
\end{equation}
where $\n R$ and $\n I_n$ are defined in Section~\ref{calibration}, $\n \EM(\boldsymbol \theta)= (\EM(\n x_i,\boldsymbol \theta),\ i=1,\ldots,n)^\top$ are evaluations of the conditional mean of the emulator and
$\n K$ is the $n\times n$  correlation matrix corresponding to the emulator, with entry $(i,j)$ given by $k((\n x_i,\boldsymbol \theta),(\n x_j,\boldsymbol\theta))$. As stated, all the parameters associated with the GaSP for the computer model are replaced by plug-in estimates. See \cite{guwangberger18} for guidance on how to obtain such estimates.

This will lead to an added computational burden in the MCMC algorithm, since now we have an additional correlation matrix $\n K$ that needs  to be recomputed for each proposal of $\boldsymbol\theta$. However, there is no additional technical or conceptual difficulty to apply the PIPS method in the context of a slow computer model.

\subsection{Discussion}\label{disc}
This section is devoted to clarify and further justify several methodological aspects of the procedure that we are proposing.
\subsubsection{One- versus two-step approach}
The idea of screening the discrepancy function for active inputs is not new: \cite{JosephYan} propose estimating the unknown parameters in \eqref{KOH} and then plugging the parameter estimates in the posterior mean of the discrepancy function $\delta$. This estimated discrepancy function is then screened for active input variables using sensitivity analysis. In this sense, their approach may be considered a \textit{two-step approach.} 

It is well-known \citep[cf. e.g. ][]{tuo2015efficient} that there are confounding issues between the vector of unknown parameters in the computer model, $\n \theta$, and the bias function itself, a topic that we will further discuss in Section~\ref{sec:confounding}. Because of this, we claim that fixing $\n \theta$ at a particular estimate is not always a sensible strategy. Instead, we propose an approach that averages over the whole posterior distribution of the unknowns and is not dependent on a single estimate of $\boldsymbol \theta$.  We illustrate the advantage of our approach with an example that we include in the Supplementary Material, Section~3. As a side-note, one might argue that our approach is indeed a two-step approach since we first obtain a sample from the posterior distribution of the parameters under the full model and then compute the posterior distribution of all competing models and resulting inclusion probabilities. That separation is merely computational and not methodological. Also, we consider this to be an advantage of our method, as the computationally heavy part is done first and the decision to screen the discrepancy function can come as an afterthought. Additionally, this allows us to  easily experiment with different values of $\alpha$ in the `spike' without having to redo the MCMC calculations.


\subsubsection{Confounding}\label{sec:confounding}

In the previous section, we mentioned the confounding issues that result from adopting of the \cite{KO2001} model paired with a GaSP prior on the discrepancy function. As a consequence, we advised against a screening strategy that hinges on a single estimate of $\n \theta$. 

Another consequence of the confounding is that, to obtain results that are authoritative from a scientific/subject-specific point of view, one should incorporate into the prior distribution of the discrepancy genuine information  about its structure, as advocated, among others, by \citet{bower2010galaxy}, \citet{vernon2010rejoinder} and \citet{Brynjarsdottir_2014}. This approach is not only difficult to implement but also at odds with the idea of producing methodology that is applicable in general. [As already mentioned, as we enforce a zero mean in \eqref{priordelta} we are already including prior information about the discrepancy function.]

We do not claim any scientific validity to the results that come out of the methodology that we are proposing: we are flagging input variables as active in the sense that, when modeling the available data using our statistical framework, the results indicate that there is considerable evidence, as measured by the posterior inclusion probability, that one needs to include a particular input variable in the discrepancy function. We have already stated the relevance of such information both to modelers and to the users of the model. 

Other authors have addressed the problem of constructing priors for the discrepancy function based of first principles, namely \cite{plumlee2017bayesian} and \cite{gu2018scaled}. These developments are however tailored to the problem of calibration. A relevant research question is to assess the impact on our posterior inclusion probabilities of adopting such priors. The discretized version of the S-GaSP of \cite{gu2018scaled} is potentially easy to accommodate.

Note that non-Bayesian alternatives to approach the calibration problem include \citet{wong2017frequentist} \citet{plumlee2019computer} and \citet{tuo2019adjustments}.

\subsubsection{Screening via range parameters}\label{range}
Recently, \cite{LinJoseph} have argued that the approach to screening that is based on a separable correlation structure like the one that we are utilizing here only allows one to assess the importance of an input variable via the magnitude of the associated range parameter --- which measures how fast the correlation along that input decays to zero---, but not in terms of how much that input variable contributes to the variability of the response. They introduce a novel correlation function which incorporates parameters that indeed (under some assumptions) measure the relative importance of each input variable in terms of how much they contribute to the variability of the response. Note that we do not claim that a larger posterior inclusion probability implies that a larger contribution to the variability of the response can be attributed to the corresponding input. In fact, we can see in Section 4 of the supplementary material that when our methodology is applied to \citeauthor{LinJoseph}'s Section 4.1 example, both inputs receive high marginal inclusion probabilities (computed as equal to $1$) while it is clear that, as far as the variability of the response is concerned, $x_1$ is more important than $x_2$. Our stance is that both inputs are indeed necessary to model the response, and that posterior inclusion probabilities are capable of quantifying the evidence in favor of this conjecture. 

A related potential issue, as pointed out by a referee, is that a variable with a linear effect on the discrepancy function may be incorrectly classified as inert. This happens because, as discussed in \cite{GL2008}, a linear (or `smooth', i.e. approximately linear, as they put it) response can be obtained in the context of a Gaussian process with large values of the range parameter, which implies a $\rho$ close to 1, and this is precisely what flags an input as inactive in our methodology. Indeed, we illustrate this phenomenon in the Supplementary Material, Section~5. Motivated by these results, in the next section we will discuss the choice of $\alpha$ guided by the ability to detect linear effects that are relevant from a practical point of view.


\subsubsection{Choice of $\alpha$}\label{choicealpha}

The hyperparameter $\alpha$ essentially determines how large should the parameter $\rho$ be in order for the associated input to be declared inert from a practical point of view. Additionally, 
as detailed in the previous section, the choice of $\alpha$ is also critical because of the fact that a high value of $\rho$ can correspond to an active input albeit one whose effect on the response is linear (or smooth, i.e., approximately linear). 

Hereafter, we provide a heuristic argument that helps determine a value for $\alpha$.
For convenience, we restrict attention to the covariance kernel defined in \eqref{rho} with $a=2$ so that the associated Gaussian process is infinitely differentiable. In this case, the derivative of the Gaussian process with respect to the input $x_\ell$ is still a Gaussian process with a variance equal to $\frac{2\sigma^2}{\psi_\ell}$. Hence, a realization of the derivative process will fall within the interval $\pm 2 \sqrt{\frac{2\sigma^2}{\psi_\ell}}$ with 95\% probability.
For an active input that has a linear effect, the derivative corresponds to the slope of the linear effect. 
We consider that if the previous bounds on the derivative are less than $20\%$ of the standard deviation of the Gaussian process then the corresponding variable can be deemed as inert from a practical point of view. This leads to $$2 \sqrt{\frac{2\sigma^2}{\psi_\ell}}<0.2\cdot \sigma \Leftrightarrow \psi_\ell > 200\Leftrightarrow \rho_\ell> \exp\left(-\frac{1}{4\cdot 200}\right)\approx 0.9987.$$
We want to choose $\alpha$ in order to discard through the spike-and-slab prior variables that, having a linear effect, that effect is very small, i.e., that have a small gradient.
Therefore, we want the spike part of the prior to have a higher density than the slab part
from this critical value (i.e., 0.9987) on. This happens when the order of magnitude of $\alpha$ is around 
5000. More precisely, we choose the value of $\alpha$ close to a critical $\alpha_0$ which is such that $\alpha_0=\arg\max\{\alpha; \mathcal{B}(0.9987\mid \alpha, 1)>\mathcal{U}(0.9987)=1\}\,.$
The choice $\alpha=5000$ is made for the simulation studies of Section~\ref{simul} and for the application of Section~\ref{sec:realapp}. As mentioned, since the computation of the PIPS is a post-processing step once an MCMC algorithm has been run, one can easily compute these quantities for different values of $\alpha$. 

More simulations about the impact of $\alpha$ are available in the Supplementary Material, Section~6.  

\subsubsection{Other correlation kernels}\label{otherck}

All the calculations in this paper are obtained for the correlation kernel \eqref{rho}, which corresponds to the power exponential kernel \eqref{pwex} using \cite{linketal2006} reparametrization $\rho_\ell=\exp(-(1/2)^a/\psi_\ell)$. This parameter is interpreted in \citet{higdonetal08} as the correlation between outputs evaluated at inputs that vary only in the $\ell$-th dimension by half their domain (recall that the inputs are scaled to vary in $[0,1]^p$). We have adopted this strategy both for simplicity and for making the comparisons in Section 7 of the Supplementary Material more transparent.
However, \citet{savetal2011} with the correlation kernel \eqref{pwex} utilize $\rho_\ell=\exp(-1/\psi_\ell)$ for $a=2$ and entertain a similar reparametrization for the Mat\'ern correlation kernel. 

\cite{guwangberger18} present a systematic parametrization of the most common correlation functions. They write all the kernels as a function of $d/\gamma$, where $d$ is the Euclidean distance and $\gamma$ is the range parameter. They then define $\rho=\exp(-1/\gamma)$ as the correlation parameter. Conceptually, this is the natural extension of our approach to other correlation kernels and we do not anticipate any reason why it should not work.

\section{Simulation studies}\label{simul}
In Section 7 of the Supplementary Material we present a simulation study devised to directly compare the results obtained using our methodology, PIPS, with those obtained using RDVS. The most important conclusion is that PIPS' results are as good as RDVS's, but PIPS is computationally less demanding. In that section, we also provide precise details regarding the MCMC scheme that we use to obtain the sample from the posterior of the unknowns under the full model.

Here, we will assess the performance of the PIPS methodology in simulated scenarios. We want to mimic situations that one might encounter when analyzing a computer model and flesh out the advantages brought by PIPS in the realm of uncertainty quantification. {These simulations can be replicated from the vignette available at
\if0\blind
{
\url{https://demiperimetre.github.io/PIPScreening/articles/ExampleScenario.html}.
}
\fi
\if1\blind
{
\textit{link removed in blind version}.
}
\fi}

With this purpose in mind, we used the values of 100 observations on 5 input variables, $\n x^\top = (x_{1}, x_{2}, x_{3}, x_{4}, x_{5})\in [0,1]^5$,  all simulated from independent uniform distributions except for $x_3$ and $x_5$, which are correlated.

We then simulated the field data according to
$$y_i = \zeta(\n x_i, \btheta) + \varepsilon_i,\quad i=1,\ldots,100,$$ 
with $\zeta(\n x_i, \btheta)$ differing in each scenario and $\varepsilon_i\sim \mathcal{N}(0,\sigma_0^2=0.05^2)$. Conceptually,   $\zeta(\n x_i, \btheta)$ is the reality which the computer model is built to replicate.

The statistical analysis of the data is constructed under the assumption that the actual  computer model is
$$\CM(\n x_i,\btheta) = \sum_{\ell=1}^3 \frac{|4 x_{\ell i} -2| + \theta_\ell }{1+\theta_\ell}\, $$ 
where $\btheta$ may either be treated as known or calibrated. We also consider a discrepancy function $\delta(\n x_i)$, which is a function of all 5 input variables, meant to capture the differences between reality and the computer model. Crucially, $\n \theta=(\theta_i,\ i=1,\ldots, 5)^\top$ but the computer model only involves the first 3 parameters.

Regarding how the computer model is related to reality, we considered three different scenarios, listed below. In all cases, $x_1$ is mishandled in the computer model. Additionally, 
  \begin{description}
 \item[Scenario 1:] $x_{2}$ has no impact in reality:  $$\zeta(\n x_i,\btheta)=\frac{|4 x_{1i}^2 -2| + \theta_1}{1+\theta_1}   +\frac{|4 x_{3i} -2| + \theta_3}{1+\theta_3}\ ;$$
 \item[Scenario 2: ] $x_{4}$ was forgotten in the computer model: $$\zeta(\n x_i,\btheta)=\frac{|4 x_{1i}^2 -2| + \theta_1}{1+\theta_1} + \frac{|4 x_{2i} -2| + \theta_2}{1+\theta_2}   +\frac{|4 x_{3i} -2| + \theta_3}{1+\theta_3} + \frac{|4 x_{4i} -2| + \theta_4}{1+\theta_4}\ ;$$
 \item[Scenario 3:] $x_{3}$ instead of $x_{5}$ appears in the computer model (and recall that, in the simulations, $x_3$ and $x_5$ are correlated): $$\zeta(\n x_i,\btheta)=\frac{|4 x_{1i}^2 -2| + \theta_1}{1+\theta_1} + \frac{|4 x_{2i} -2| + \theta_2}{1+\theta_2}   +\frac{|4 x_{5i} -2| + \theta_5}{1+\theta_5}\ .$$
\end{description}

In the simulations, the vector of model parameters is chosen as 
$(\theta_1,\theta_2,\theta_3,\theta_4,\theta_5)^\top=(0.4,0.5,0.6,0.7,0.8)^\top$. MCMC and prior specifications were  detailed in the Supplementary Material, Section 7.

 \begin{table}
\centering
\scriptsize{
\begin{tabular}{llaarrr}
 \hline
 Sc. 1&& $x_1$ & $x_2$ &$x_3$ & $x_4$ & $x_5$  \\ 
  \hline
\multirow{3}{*}{Fixed}&th0.1&1.00& 1.00& 0.00& 0.00& 0.00 \\ 
 & th0.5  &1.00& 1.00& 0.00& 0.00& 0.00\\ 
  &th09   &1.00& 1.00& 0.00& 0.00& 0.00\\ 
  \hline
  \multirow{3}{*}{Calibrated}&th0.1 &1.00& 1.00& 0.00& 0.00& 0.00\\ 
  &th0.5 &1.00& 1.00& 0.00& 0.00& 0.00 \\ 
  &th0.9&1.00& 1.00& 0.00& 0.00& 0.00 \\ 
  \hline
\end{tabular}
\begin{tabular}{llarrar}
 \hline
 Sc. 2&& $x_1$ & $x_2$ &$x_3$ & $x_4$ & $x_5$  \\ 
  \hline
\multirow{3}{*}{Fixed}&th0.1&1.00& 0.00& 0.00& 1.00& 0.00 \\ 
 & th0.5  &1.00& 0.00& 0.00& 1.00& 0.00\\ 
  &th09   &1.00& 0.00& 0.00& 1.00& 0.00\\ 
  \hline
  \multirow{3}{*}{Calibrated}&th0.1 &1.00& 0.00& 0.00& 1.00& 0.00\\ 
  &th0.5 &1.00& 0.00& 0.00& 1.00& 0.00 \\ 
  &th0.9&1.00& 0.00& 0.00& 1.00& 0.00 \\ 
  \hline
\end{tabular}

\begin{tabular}{llarara}
 \hline
 Sc. 3&& $x_1$ & $x_2$ &$x_3$ & $x_4$ & $x_5$  \\ 
  \hline
\multirow{3}{*}{Fixed}&th0.1&1.00& 0.00& 1.00& 0.00& 1.00 \\ 
 & th0.5  &1.00& 0.00& 1.00& 0.00& 1.00\\ 
  &th09   &1.00& 0.00& 0.98& 0.00& 1.00\\ 
  \hline
  \multirow{3}{*}{Calibrated}&th0.1 &1.00& 0.00& 1.00& 0.00& 1.00\\ 
  &th0.5 &1.00& 0.00& 1.00& 0.00& 1.00  \\ 
  &th0.9&1.00& 0.00& 0.97& 0.00& 1.00 \\ 
  \hline
\end{tabular}}
\caption{Proportion of simulations where the PIPS are above the thresholds $th=0.1,0.5,0.9$ for the three scenarios (Sc.). The parameter $\btheta$ can be either fixed or calibrated. Shadowed columns represent the truly active variables in the discrepancy function.} 
\label{tab:withoutcal}
 \end{table}

Table \ref{tab:withoutcal} summarizes the results of the simulation under the different scenarios by reporting the proportion of times that the posterior inclusion probabilities of the input variables are above three different thresholds: 0.1, 0.5 and 0.9. 
The method is able to clearly separate  the inert variables from the active ones: the PIPS of inert variables are all below 0.1 and the PIPS of active variables are almost always above 0.9 --- notice how scenario 3 is slightly more challenging.

\section{A photovoltaic plant computer model}\label{sec:realapp}

In this section, we introduce a real application which was originally studied by \citet{carmassiPV}. The idea is to use a computer model to predict the power produced by a photovoltaic plant (PVP)  given some meteorological conditions such as the temperature and the sun irradiation. The computer model also depends on some parameters which can be fixed at so-called nominal values specified using expert knowledge or can be calibrated from experimental data. In \citet{carmassiPV}, the authors showed that the discrepancy term plays an important role in the statistical model but they did not investigate whether all the input variables were active. 
 
Here, we will use the experimental data in order to learn about possible model misspecification by screening the discrepancy function using the PIPS methodology.  
The computer model  aims at mimicking the behavior of a PVP consisting of 12 panels connected together. This computer model is fast to run and can be represented by a function $\CM:\mathbb{R}^4\times\mathbb{R}^6\owns(\n x,\btheta)\mapsto y\in\mathbb{R}$. The input variables $\n x=(t ,I_g, I_d, T_e)^\top$ consist of $t$, the UTC time since the beginning of the year, $I_g$, the global irradiation of the sun, $I_d$ the diffuse irradiation of the sun, and $T_e$, the ambient temperature. The parameter $\btheta$ consists actually of $6$ parameters but only one parameter is calibrated, the module photo-conversion efficiency. A sensitivity analysis has proven the other parameters to be of negligible importance. The output $y$ is the resulting instantaneous power delivered by the PVP.

The experimental data correspond to the test stand of 12 panels and were collected over two months every $10$ seconds, which makes for a huge amount of data. Only the data for which the production is positive are kept, corresponding  to daytime measurements.
The data contain the power production, which is to be compared with the output of the computer model; the four input variables of the computer model and an additional temperature (temperature on the panel, $T_p$) were also recorded. Although this last variable is not an input variable of the computer model, it is tested as a potential active variable in the discrepancy.

 In order to detect the input variables which may cause the discrepancy between the computer model output and reality, 
 we independently applied the PIPS methodology to the data obtained on each of several days. The considered days are those in August and September 2014. September 7th was removed since the recorded production was null which is a consequence of a sensor malfunction; hence, the methodology was applied independently to a total of sixty days.
 
\begin{figure}
  \centering
  \begin{tabular}{cc}
  \includegraphics[scale=.3]{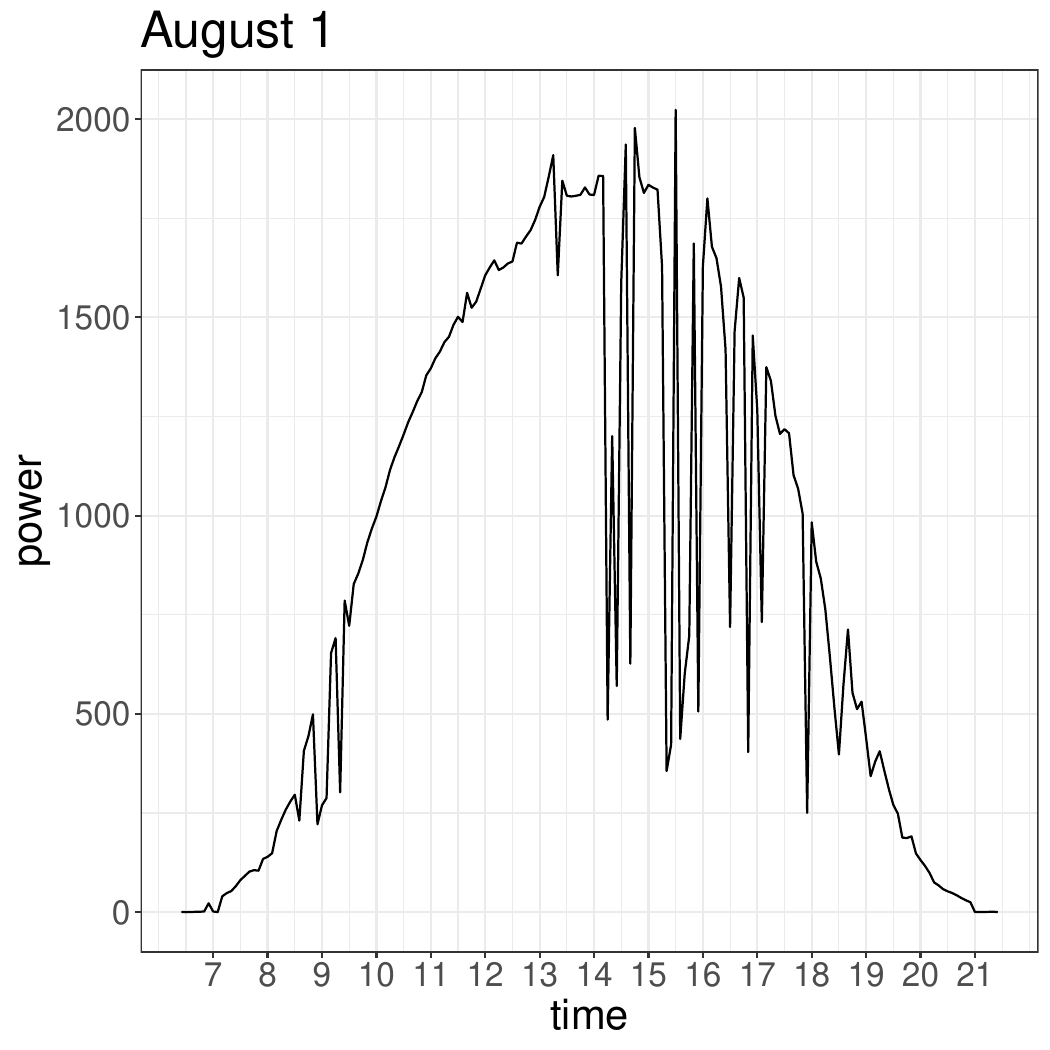} & \includegraphics[scale=.3]{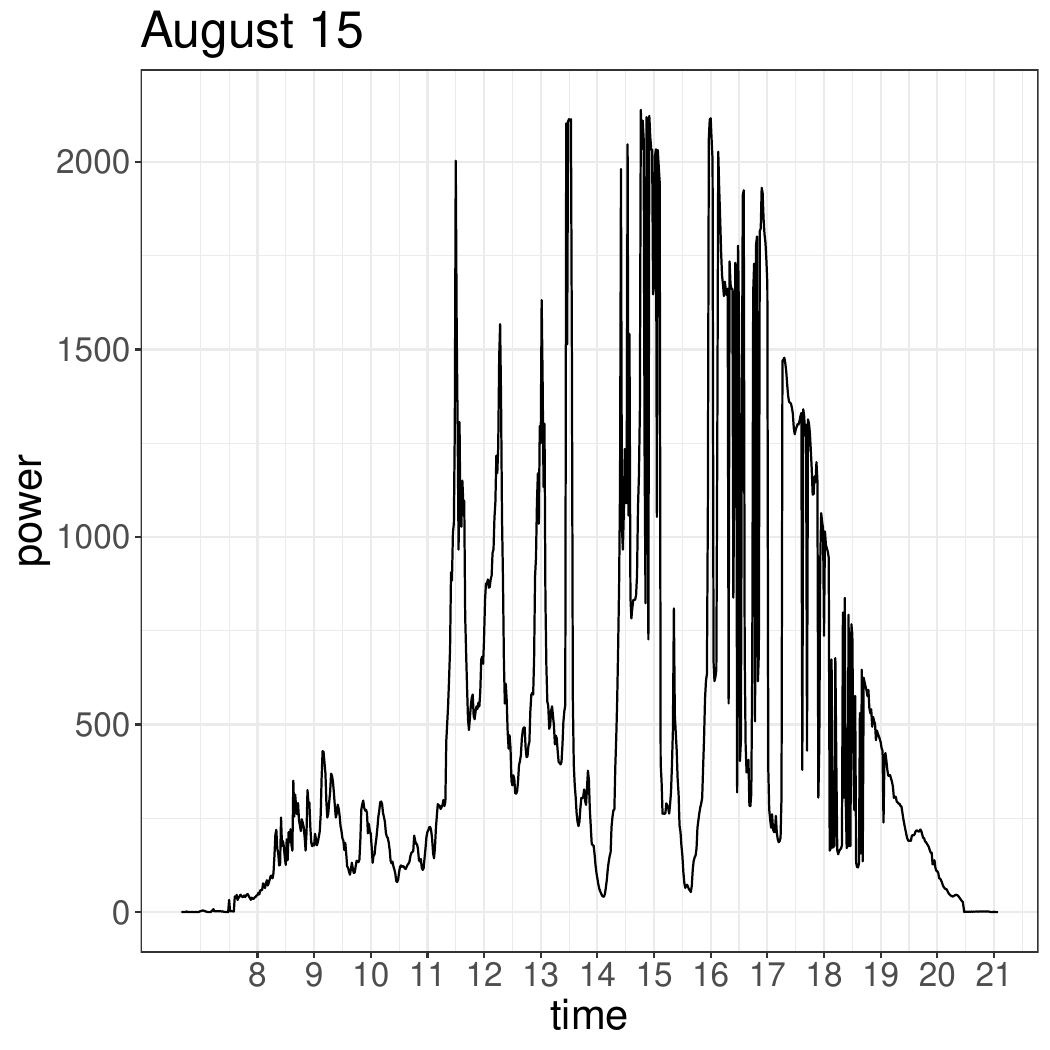}\\
  \includegraphics[scale=.3]{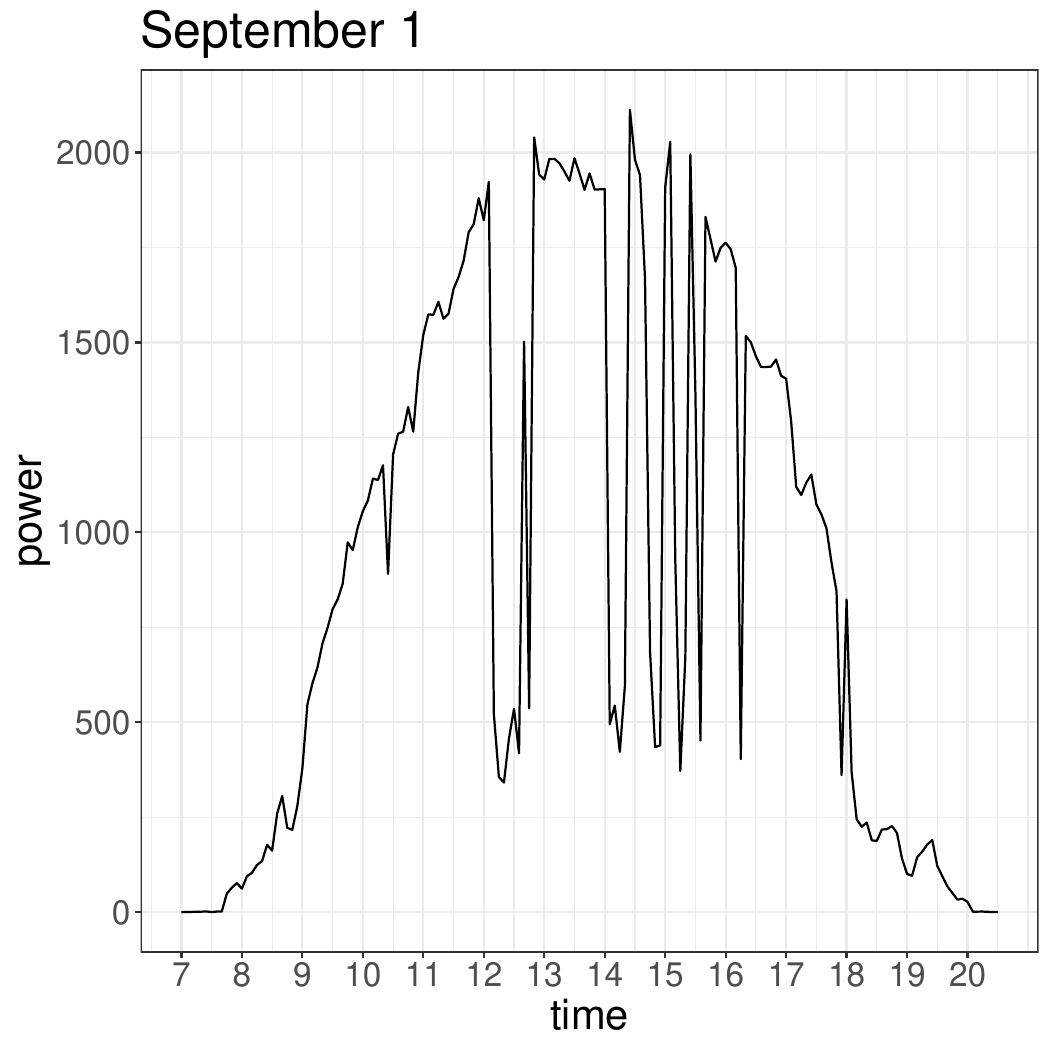} & \includegraphics[scale=.3]{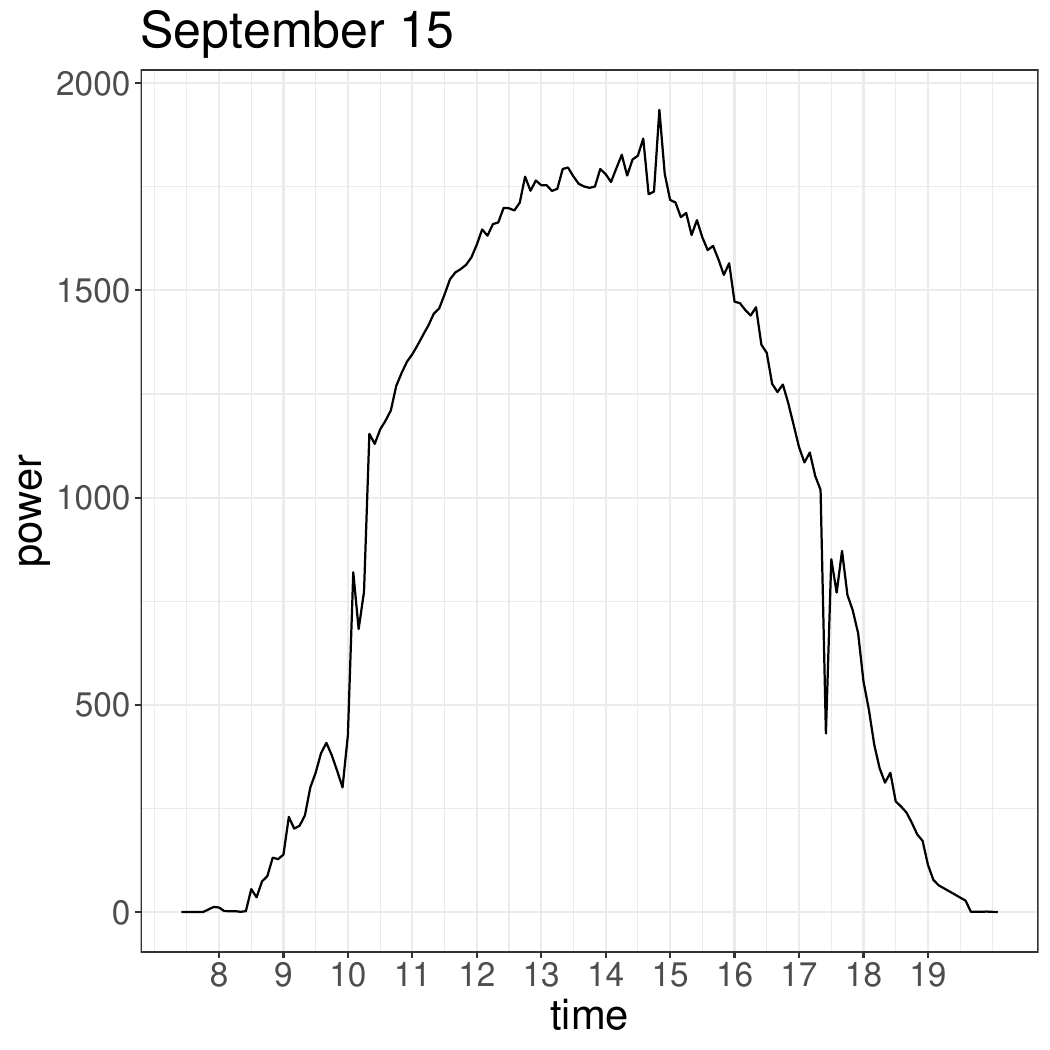}\\
  \end{tabular}
  \caption{Power produced by the PV power plant on days: August 1st, August 15th, September 1st, September 15th. A measure every five minutes is kept for a positive production.}
  \label{figPVpower}
 \end{figure}

 To limit the computational burden for a given day, we took a measurement every five minutes, 
 which gives between 99 and 178 data points per day. Figure \ref{figPVpower} illustrates four different days of power recording.
 Although these four days exhibit somewhat different patterns of production, the common feature is the global increase in production beginning at sunrise followed by a decay which reaches a null production when the sun sets. Higher frequency peaks may be a consequence of clouds limiting the sun irradiation. We emphasize that this case study is challenging because of the structure of the correlation between the input variables in the computer model. Since a measurement was taken every five minutes, we assume that the correlation that appears in the outputs is simply the result of the correlation within the inputs, which is consistent with the independence assumption on the error terms in \eqref{KOH}.
 
We consider the situation where all the parameters were fixed at their nominal value and when 
the  module photo-conversion efficiency calibration and the detection of active variables are performed simultaneously. All the input variables and the model parameter were scaled to the $[0,1]$ interval. The output data
were normalized by subtracting their mean and dividing by their standard deviation.
The prior distributions for the variances of the measurement error and the discrepancy
were chosen as inverse gamma distribution with parameters favoring a larger variance for the discrepancy and in line with the known precision of sensors,
namely, $\mathcal{IG}(4,1/400)$ for the measurement error and $\mathcal{IG}(2,1/200)$ for the variance corresponding to the GaSP which models the discrepancy.
A uniform prior within an interval provided by experts was used for the only parameter to calibrate. We also choose $a=1.9$ in the correlation kernel defined in \eqref{rho}, and set $\alpha=5000$. 
The MCMC set-up is as detailed in the Supplementary Material, Section 7. 

\begin{figure}
  \centering
  \begin{tabular}{c}
   \includegraphics[scale=.5]{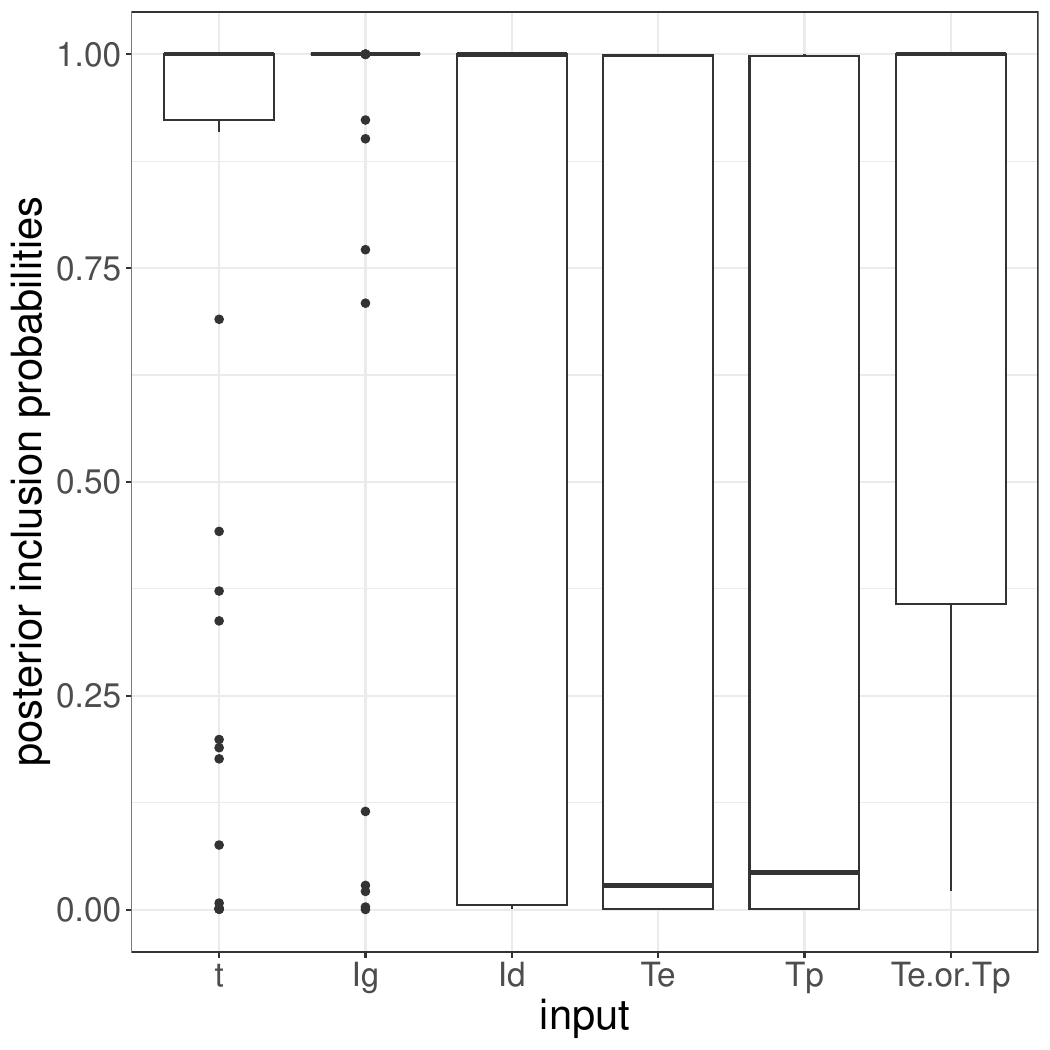}\\
   \includegraphics[scale=.5]{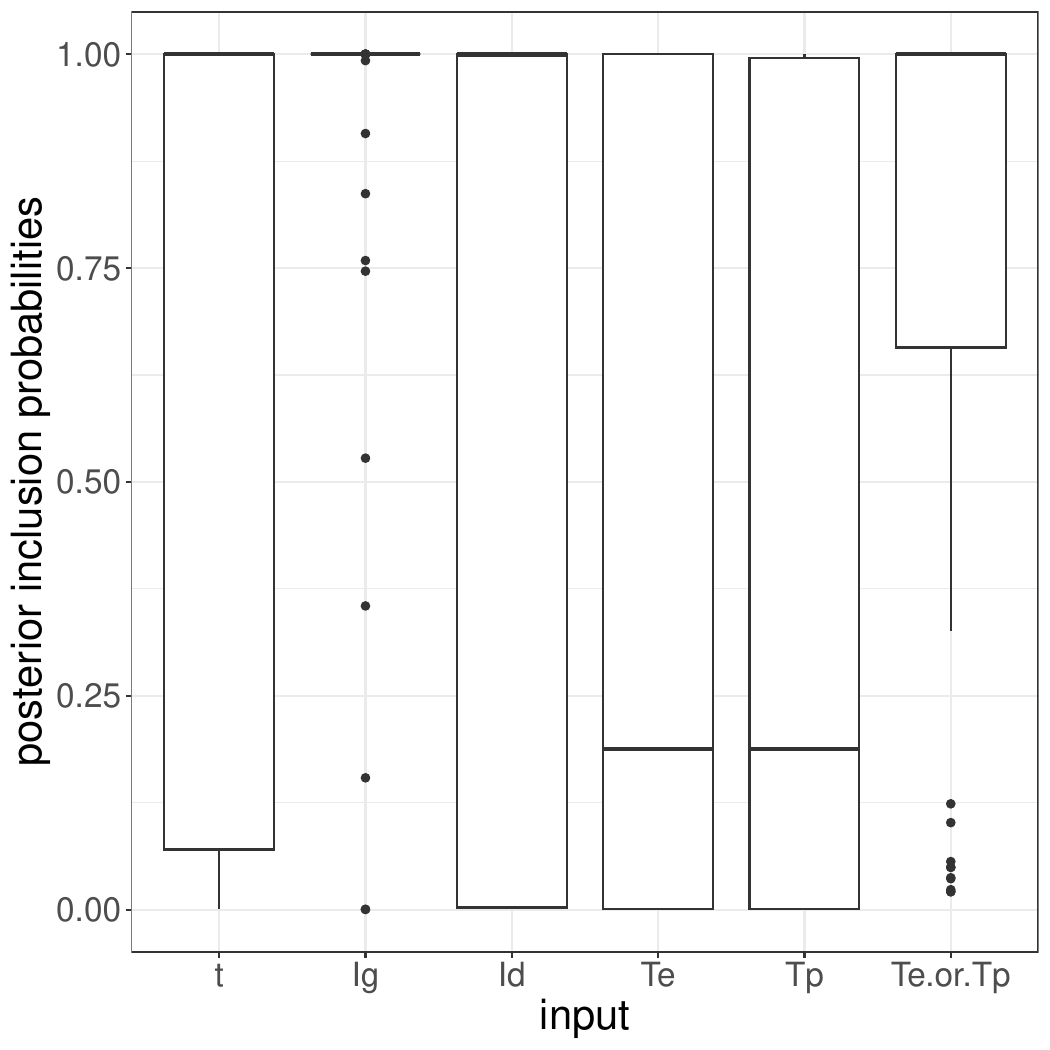}\\
  \end{tabular}
\caption{Boxplots of the posterior inclusion probabilities for the input variables in the discrepancy computed for the 60 days of data. The column $(T_e \textrm{ or } T_p)$ corresponds to the probability that at least one of the two temperatures is active.
The top boxplots correspond to the case when the parameter $\theta$ is fixed at its nominal value. The bottom boxplots correspond to the case when $\theta$ is calibrated.} 
\label{figboxplotPV}
 \end{figure}

Using the resulting sample, 
we obtained the posterior inclusion probabilities for all the inputs on each day as well as the posterior probability 
that, on each day, at least one of the two temperatures ($T_e$ and $T_p$) is active in the discrepancy. Figure \ref{figboxplotPV} provides the boxplots of these probabilities computed for the sixty days of data, that is, the boxplots describe the set of sixty posterior inclusion probabilities, each corresponding to one of 60 the days. 
The posterior inclusion probabilities are summarized over various conditions that are encountered during the sixty days, and can be thought as providing a sample from a posterior predictive distribution. 

The results indicate that  some inputs are active for almost all conditions while for some others some conditions lead to inert variables. 
For the two first variables (time $t$ and global irradiation $I_g$) the results are clearly in favor of including them in the discrepancy for any kind of conditions. For the diffuse irradiation $I_d$ and the two temperatures the results are less clear, although they seem to indicate that at least one of the two temperatures should be included in the discrepancy for most of the sixty days. 
The variability observed in the inclusion probabilities could indicate that, for some conditions, the corresponding input variables are correctly handled by the computer model while, for 
 other conditions, they should be included in the discrepancy. 
 It could be of interest for the scientists who developed the model to investigate in depth what are the conditions that make these variables active or inert in the discrepancy.
 When the computer model is simultaneously calibrated, the calibrated value is shifted toward its upper bound (more than $75\%$ of the days) but the inclusion probabilities are quite similar. 
 
\section{Conclusion}\label{conclusion}

We focused on the problem of screening the discrepancy function, which is an object used to account for the gap between a real phenomenon and its ``in silico'' simulation. The outcome of this screening procedure provides the practitioner and the modeler with valuable information regarding potential flaws of the computer model by determining which input variables are active in the discrepancy function.
We cast this screening procedure into the more general problem of variable selection for GaSP regression. We then proposed an original method, named PIPS, which aims to select the active variables on the basis of a single MCMC sample.
The efficiency of the PIPS method was demonstrated on synthetic examples and was also illustrated on a real world application.
Although the computer models in the applications were fast enough so that we did not need to resort to an emulation step, the PIPS method could be combined with an emulator.
However, one has to be careful that, in this case, the discrepancy may account for both the flaws of the computer model and for the inaccuracies of its emulator.


PIPS is not quite an automatic procedure because of the steps that are necessary to obtain an MCMC sample from the posterior distribution of the unknowns under the full model, making sure that the algorithm is mixing properly. An important improvement would be to devise an adaptive MCMC scheme that could safely be deployed to obtain the sample automatically. Such an adaptive MCMC scheme could be combined with some intermediate results from PIPS in order to reinforce the sampling in the dimensions where the input variables are most likely to be active. A complementary issue is the determination of the area of validity of the computer model. PIPS could be used to feed adaptive design strategies to accurately determine this region.



\section*{Supplementary Materials}

\textbf{PDF Supplement:} This supplement contains seven sections that correspond to (1) the proof of Theorem 1, (2) technical details about the derivation of the PIPs estimator, (3) a simulation study to compare a one-step approach versus a two-step approach, (4) the application of the PIPS approach to the Lin and Joseph's Section 4.1 example, (5) a simulation study in a linear scenario, (6) a simulation study to investigate the effect of the choice of the parameter $\alpha$ and (7) a comparison with the RDVS method.

\noindent \textbf{ZIP file supplement:} This supplement contains the R package PIPScreening with R files for reproducing some of the simulation studies presented in the paper or in the PDF supplement. A README file is provided to connect the files with the corresponding simulation studies.

\if0\blind
{

\section*{Acknowledgments}
We greatly thank the Editors and the reviewers whom provided insightful comments that helped us improve the early versions of the paper.
%
}\fi

\section*{Funding}
This work was partially developed while the authors were visiting the Statistical and Applied Mathematical Sciences Institute, and were hence partially supported by the National Science Foundation under Grant DMS-1638521.
This paper has been partially supported by research grants PID2019-104790GB-I00 funded by MCIN/AEI/ 10.13039/501100011033 and PID2022-138201NB-I00 funded by MCIN/AEI /10.13039/501100011033 / FEDER, UE Una manera de hacer Europa”. 
Pierre Barbillon has received support from the Marie-Curie FP7 COFUND People
Programme of the European Union, through the award of an AgreenSkills/AgreenSkills+ fellowship (under grant agreement n$\circ$609398).
 Rui Paulo was partially supported by the Project CEMAPRE/REM---UIDB/05069/2020---financed by FCT/MCTES through national funds, and by the sabbatical fellowship SFRH/BSAB/142992/2018 attributed by FCT. 

\section*{Disclosure statement}
The authors report there are no competing interests to declare.

\textsl{\bibliographystyle{chicago}
\bibliography{biblio}}

\end{document}



\def\spacingset#1{\renewcommand{\baselinestretch}%
{#1}\small\normalsize} \spacingset{1}


\if0\blind
{
  \title{\bf Supplementary Material to \\ Screening the Discrepancy Function of a Computer Model}
  \author{Pierre Barbillon\hspace{.2cm}\\
    Université Paris-Saclay, AgroParisTech, INRAE, UMR MIA Paris-Saclay,\\ 91120, Palaiseau, France\\
    and \\
    Anabel Forte \\
    Department of Statistics and Operations Research, Universitat de Valencia\\
    and
    \\
    Rui Paulo \\
    CEMAPRE/REM and Department of Mathematics, \\ Lisbon School of Economics and Management, Universidade de Lisboa
    }
  \maketitle
} \fi

\if1\blind
{
  \bigskip
  \bigskip
  \bigskip
  \begin{center}
    {\LARGE\bf Supplementary Material to \\ Screening the Discrepancy Function of a Computer Model}
\end{center}
  \medskip
} \fi

\section{Proof of  Theorem 1 }\label{proof}

\paragraph{Conditions on the prior distribution}
We assume that the prior distribution $\pi(\n \eta)=\pi(\n \theta,\sigma^2,\sigma_0^2)$ has a compact support that does not contain points such that $\sigma^2=0$ or $\sigma_0^2=0$. Moreover, we assume this prior to be proper and bounded.

\paragraph{Proof}
Without loss of generality, let us assume that $\ell=1$. 
We consider the sequence of prior distributions $$\pi^\alpha(\rho_1)=\mathcal{B}(\rho_1\mid \alpha, 1)= \alpha \rho_1^{\alpha-1}; \; 0<\rho_1<1.$$ Let $\n \eta= (\sigma^2,\sigma_0^2, \n \theta)^\top$ and $\n \rho_{(-1)}=(\rho_2,\ldots,\rho_p)^\top$. Additionally, denote by $g(\boldsymbol y \mid \brho,\bneta)$ the density of the field data corresponding to the sampling distribution given in equation (5) of the main document. 

\textbf{First step:}
we prove that, as a consequence of the dominated convergence theorem,
\begin{align*}
& \lim_{\alpha \rightarrow \infty} \int_{\bneta,\brho_{(-1)}}\int_{\rho_1} g(\boldsymbol y\mid \brho,\bneta)\ \pi(\brho\mid \n \gamma)\ \pi(\bneta)\ d\rho_1\ d \bneta\ d\brho_{(-1)}\\
&=\int_{\bneta,\brho_{(-1)}}\left(\lim_{\alpha\rightarrow \infty}\int_{\rho_1} g(\boldsymbol y \mid \brho,\bneta)\ \pi^\alpha(\rho_1)\ d\rho_1\right)\pi (\brho_{(-1)}) \ \pi(\bneta)\ d\bneta\ d\brho_{(-1)}\ .
\end{align*}

The dominated convergence theorem applies, since for all $\alpha$, we have
$$\int_{\rho_1} g(\boldsymbol y\mid \brho,\bneta)\ \pi^\alpha(\rho_1)\ d\rho_1\le \sup_{\brho,\bneta} g(\boldsymbol y\mid \brho,\bneta)\int \pi^\alpha(\rho_1)\ d\rho_1 =  \sup_{\brho,\bneta}\ g(\boldsymbol y\mid \brho,\bneta)\cdot 1 <+\infty \,.$$
Indeed, the inequality above ensures that
the function $$\brho_{(-1)},\bneta\mapsto \int_{\rho_1} g(\boldsymbol y\mid \brho,\bneta)\ \pi^\alpha(\rho_1)\ d\rho_1\ \pi(\brho_{(-1)})\ \pi(\bneta)$$ is bounded, for all $\alpha$, by $\sup_{\brho,\bneta} g(\boldsymbol y\mid \brho,\bneta)\ \pi(\brho_{(-1)})\ \pi(\bneta) $ which is integrable provided that the supremum is finite. 
The supremum is taken over the compact set corresponding to the compact support of the prior distribution of $\bneta$ and $\brho$. This supremum
is finite since the priors are bounded and 
$\sup_{\brho,\bneta} g(\boldsymbol y\mid \brho,\bneta)$ is the supremum of the image of a compact set by a continuous function (a Gaussian likelihood as a function of its parameters where there are no singularities because the compact domain does not contain configurations of parameters such that $\sigma^2=0$ or $\sigma_0^2=0$).
Therefore, the dominated convergence theorem applies.\\

\textbf{Second step:}
By using integration by parts, we have 
\begin{align*}
&\int_{\rho_1}g(\boldsymbol y\mid \brho,\bneta)\ \pi^\alpha(\rho_1)d\rho_1=\int_{\rho_1}g(\boldsymbol y\mid \brho,\bneta)\ n  \rho_1^{n-1}d\rho_1\\
&=\left[g(\boldsymbol y\mid \brho,\bneta)\right]_{\rho_1=0}^{\rho_1=1}-\int \frac{\partial }{\partial\rho_1}\ g(\boldsymbol y\mid \brho,\bneta)\ \rho_1^{\alpha}d\rho_1\\
&=g(\boldsymbol y\mid (1,\brho_{(-1)}),\bneta)\cdot 1 - g(\boldsymbol y\mid (0,\brho_{(-1)})),\bneta)\cdot 0 -\int \frac{\partial }{\partial\rho_1}g(\boldsymbol y\mid \brho,\bneta)\ \rho_1^{\alpha}\ d\rho_1\ .
\end{align*}

\textbf{Third step:} by showing that $\int \frac{\partial }{\partial\rho_1}\ g(\boldsymbol y\mid \brho,\bneta)\ \rho_1^{\alpha}\ d\rho_1\rightarrow 0$ as $\alpha\rightarrow\infty$, the proof will be complete.

We apply again the dominated convergence theorem. For $\alpha_0$ large enough, we have:
\begin{equation}
 \nonumber
 \left|\frac{\partial }{\partial\rho_1} g(\boldsymbol y\mid \brho,\bneta)\ \rho_1^{\alpha}\right|\le \left|\frac{\partial }{\partial\rho_1}g(\boldsymbol y\mid \brho,\bneta)\ \rho_1^{\alpha_0}\right|\,.
\end{equation}
The function $\rho_1\mapsto \frac{\partial }{\partial\rho_1}\ g(\boldsymbol y\mid \brho,\bneta)$ is continuous in $[0,1]$ and $C^\infty$ in $(0,1]$ because of the form of the covariance function and the polynomial expression of the inverse and the determinant functions in the likelihood.
We can extend by continuity to $0$ the function $\rho_1\mapsto \frac{\partial }{\partial\rho_1}\ g(\boldsymbol y\mid \brho,\bneta)\ \rho_1^{\alpha_0}$, the evaluation of which will be $0$ since the derivative of the likelihood is polynomial in $1/\rho_1$.
For this reason, $\alpha_0$ needs to be chosen large enough to compensate.

\section{The PIPS estimator}\label{IS} 

Let $\n \nu =(\n \eta,\n \rho)$. Then, Bayes theorem establishes that
$$
\pi(\n \nu\mid \n y,\n \gamma) = \frac{g(\n y\mid \n \nu,\n \gamma)\ \pi(\n \nu\mid \n \gamma)}{m(\n y\mid \n \gamma)}
\Leftrightarrow g(\n y\mid \n \nu,\n \gamma)\ \pi(\n \nu\mid \n \gamma) =m(\n y\mid \n \gamma)\ \pi(\n \nu\mid \n y,\n \gamma)
$$
so that the right-hand side of Equation (14) of the paper reduces to
\begin{align}
\int & \frac{m(\n y\mid \n \gamma)\ \pi(\n \nu\mid \n y,\n \gamma)}{m(\n y\mid \n \gamma =\n 1)\ \pi(\n \nu\mid \n y,\n \gamma =\n 1)}\ \pi(\n \nu\mid \n y,\n \gamma =\n 1)\ d\n \nu \notag \\
&\quad = \int B_{\boldsymbol \gamma}\ \frac{\pi(\n \nu\mid \n y,\n \gamma)}{\pi(\n \nu\mid \n y,\n \gamma =\n 1)}\ \pi(\n \nu\mid \n y,\n \gamma =\n 1) \ d\n \nu \label{ISW}\\
& \quad = B_{\boldsymbol \gamma}\ \int \pi(\n \nu\mid \n y,\n \gamma)\ d\n \nu \notag\\
& \quad = B_{\boldsymbol \gamma} \notag
\end{align}
which proves the result given in this equation and then Equation (15) of the paper provides  
 an estimator of $B_{\boldsymbol \gamma}$. Expression \eqref{ISW} makes it clear that we are in the presence of an importance sampling estimator with importance weights 
$$
\frac{\pi(\n \nu\mid \n y,\n \gamma)}{\pi(\n \nu\mid \n y,\n \gamma =\n 1)}= B_{\boldsymbol \gamma}^{-1} \
\pi(\n \rho \mid \n \gamma) = B_{\boldsymbol \gamma}^{-1} \prod_{\ell:\ \gamma_{\ell}=0} \alpha_{\ell}\ \rho_{\ell}^{\alpha_\ell-1}\leq B_{\boldsymbol \gamma}^{-1} \alpha^{p-\boldsymbol 1'\boldsymbol \gamma}\leq B_{\boldsymbol \gamma}^{-1} \alpha^p
$$
because we set $\alpha_{\ell}=\alpha>1$. As a consequence, the variance of the proposed estimator is finite, since
\begin{align*}
\int & B_{\boldsymbol \gamma}^2\ \frac{\pi(\n \nu\mid \n y,\n \gamma)^2}{\pi(\n \nu\mid \n y,\n \gamma =\n 1)^2}\ \pi(\n \nu\mid \n y,\n \gamma =\n 1) \ d\n \nu \\
&\quad =\int  B_{\boldsymbol \gamma}^2\ \frac{\pi(\n \nu\mid \n y,\n \gamma)}{\pi(\n \nu\mid \n y,\n \gamma =\n 1)}\ \pi(\n \nu\mid \n y,\n \gamma) \ d\n \nu  \\
&\quad \leq B_{\boldsymbol \gamma}^2\ \frac{\alpha^p}{B_{\boldsymbol \gamma}}\ \int \pi(\n \nu\mid \n y,\n \gamma) \ d\n \nu\\
& \quad =  B_{\boldsymbol \gamma}\ \alpha^p\ .
\end{align*}

\section{Comparison of one- versus two-step approach.}

We aim to motivate the single step approach with a simple example.
We consider the true physical model as: 
$$\zeta_i(x_1,x_2)=x_1+x_2,.$$

The field data are simulated as, for $i=1,\ldots,n$ :
$$y_i=\zeta(\n x_i)+\epsilon_i\,,$$

where $\epsilon_i\overset{iid}{\sim}\mathcal{N}(0,\sigma^2_\epsilon=0.05^2)$ and $\n x_i$ are drawn uniformly in $[0,10]$.

The available computer model is 
$$f(\mathbf{x},{\theta})= \theta x_1+(1-\theta)x_2\,.$$
 
 The parameter ${\theta}$ is a model parameter, it may be fixed or  to calibrate. This parameter tunes which of the input variables $x_1$ or $x_2$ is taken into account with the constraint that they cannot  both be active.
 
 We compare four strategies to compute the posterior inclusion probabilities for $x_1$ and $x_2$: i) nominal: fix $\theta=0.5$, ii) mean: run a calibration step, fix $\theta$ to the mean of its posterior distribution, iii) median: run a calibration step, fix $\theta$ to the median of its posterior distribution, iv) calibrate and use the posterior sample to compute the posterior inclusion probabilities.
 Strategies ii) and iii) are two-step approaches while strategy iv) is a single step approach.

\begin{figure}
\centering
 \includegraphics[]{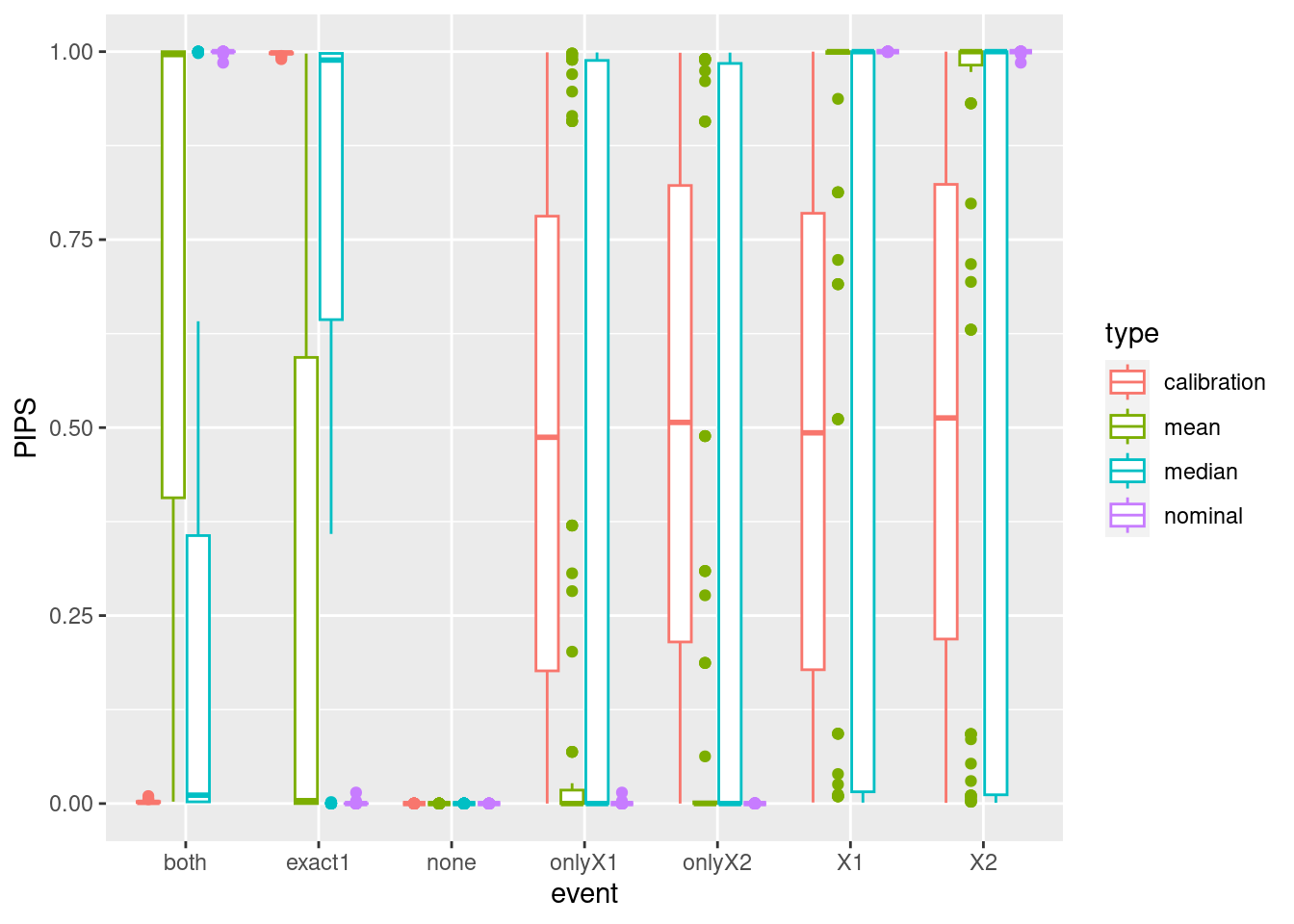}
 \label{fig1vs2}
 \caption{Computed posterior inclusion probabilities over 100 repetitions for the four different strategies: i)nominal, ii)mean, iii) median, iv) calibration for the possible inclusion in the discrepancy: both input variables included, exactly one of them, none of them, only $x_1$, only $x_2$, $x_1$, $x_2$.}
\end{figure}

We repeat the simulation settings a hundred times and plot the posterior inclusion probabilities related to the input variables that should be included in the discrepancy. Figure \ref{fig1vs2} displays the computed posterior probabilities of the different possible events. The specificity of this example lies in the 
bimodal distribution for $\theta$. The different possible choice for fixing the calibration parameter (the mean or the median of its posterior distribution e.g.) may lead to contrasted analysis of the discrepancy contrary to the single step approach which provides a more stable response. Indeed, it detects that 
$x_1$ and $x_2$ have posterior inclusion probabilities around $0.5$ and that they cannot be included together which is consistent with the case at hand.

The R code to reproduce this example is provided in \texttt{SI1.html}.

\section{Application of PIPS to the Lin and Joseph's Section 4.1 example}

We test our method on the toy example provided in Section 4.1 of \citet{LinJoseph}.
We compute $n=20$ outputs from this model:
$$y(\n x) = \exp (2 \sin(0.5\pi x_1 ) + 0.5 \cos(2.5\pi x_2 )).$$
We compute the PIPS on these data as if the $\n y$ were the residuals computed between the field experiments and the outputs of a simulator.
With the same settings as in the rest of the paper, we retrieve PIPS being equal to $1$ for both inputs $x_1$ and $x_2$ despite  the different smoothness with respect to these two inputs.

The R code to reproduce this example is provided in \texttt{SI2.html}.


\section{Performance in a linear effect scenario}

We consider a linear model with decreasing effects:
$$\zeta(x_1,\ldots,x_5)=2\cdot x_1 + 1\cdot x_2+.5\cdot x_3 + .25\cdot x_4+ 0\cdot x_5 \,.$$
An additional variable $x_5$  that has no effect at all is added.
The field data are for $i=1,\ldots,n$:
$$y_i=\zeta(\n x_i)+\epsilon_i\,,$$
where $\epsilon_i\overset{iid}{\sim}\mathcal{N}(0,\sigma^2_\epsilon=0.25^2)$ and $\n x$ being simulated uniformly in $[0,1]^5$.
As in the previous section, we consider these data as the residuals between the field data and the outputs of the computer model.  This can be regarded as screening a computer model as in \citet{linketal2006}.

\begin{figure}
\centering
 \includegraphics[scale=.5]{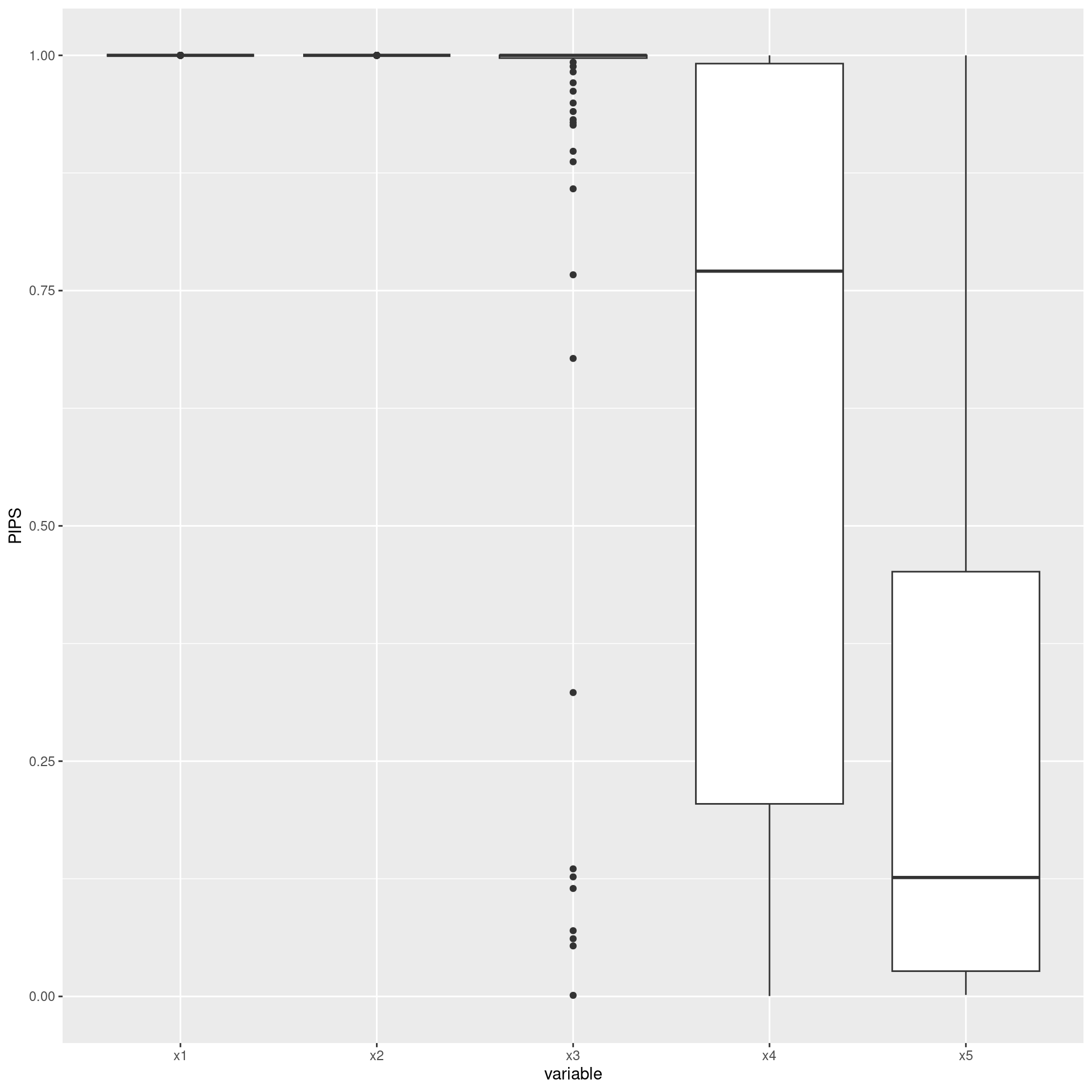}
 \caption{Posterior inclusion probabilities computer over 100 repetitions for the five input variables.}
  \label{figlin}
\end{figure}

Figure \ref{figlin} displays the computed inclusion probabilities over 100 repetitions of the setting described above. 
These probabilities allow us to detect active variables even if they have a linear effect provided that this effect is strong enough.

The R code to reproduce this example is provided in \texttt{SI3.html}.

\section{The importance of the choice of $\alpha$}

In order to assess the effect of the parameter $\alpha$ on the computation of 
the posterior inclusion probabilities we simulate data according to the following setting.

The true physical model is 
$$\zeta(x_1,\ldots,x_7)=\frac{|4 x_{1}^2 -2| + \theta_1}{1+\theta_1}   +\frac{|4 x_{2} -2| + \theta_2}{1+\theta_2}+\frac{|4 x_{3} -2| + \theta_3}{1+\theta_3} + 4\cdot\theta_4\cdot x_4  +  2\cdot\theta_5\cdot x_5 + \theta_6\cdot x_6 \,.$$

The field data are for $i=1,\ldots,N$ :
$$y_i=\zeta(\mathbf{x_i})+\epsilon_i\,$$

where $\epsilon_i\overset{iid}{\sim}\mathcal{N}(0,\sigma^2_\epsilon=0.05^2)$ and $\n x_i$ being drawn uniformly in $[0,1]^7$.

Our computer model is 
$$f(\mathbf{x},\boldsymbol{\theta})= \frac{|4 x_{1} -2| + \theta_1}{1+\theta_1} +\frac{|4 x_{2} -2| + \theta_2}{1+\theta_2}   \,.$$
Thus, $x_1$ is incorrectly taken into account in $f$ and $x_3$ is "forgotten" as well as all the linear terms  in the computer model. The parameter $\boldsymbol{\theta}$ is a model parameter, it may be fixed or calibrated.

\begin{figure}
 \centering
 \includegraphics[scale=.6]{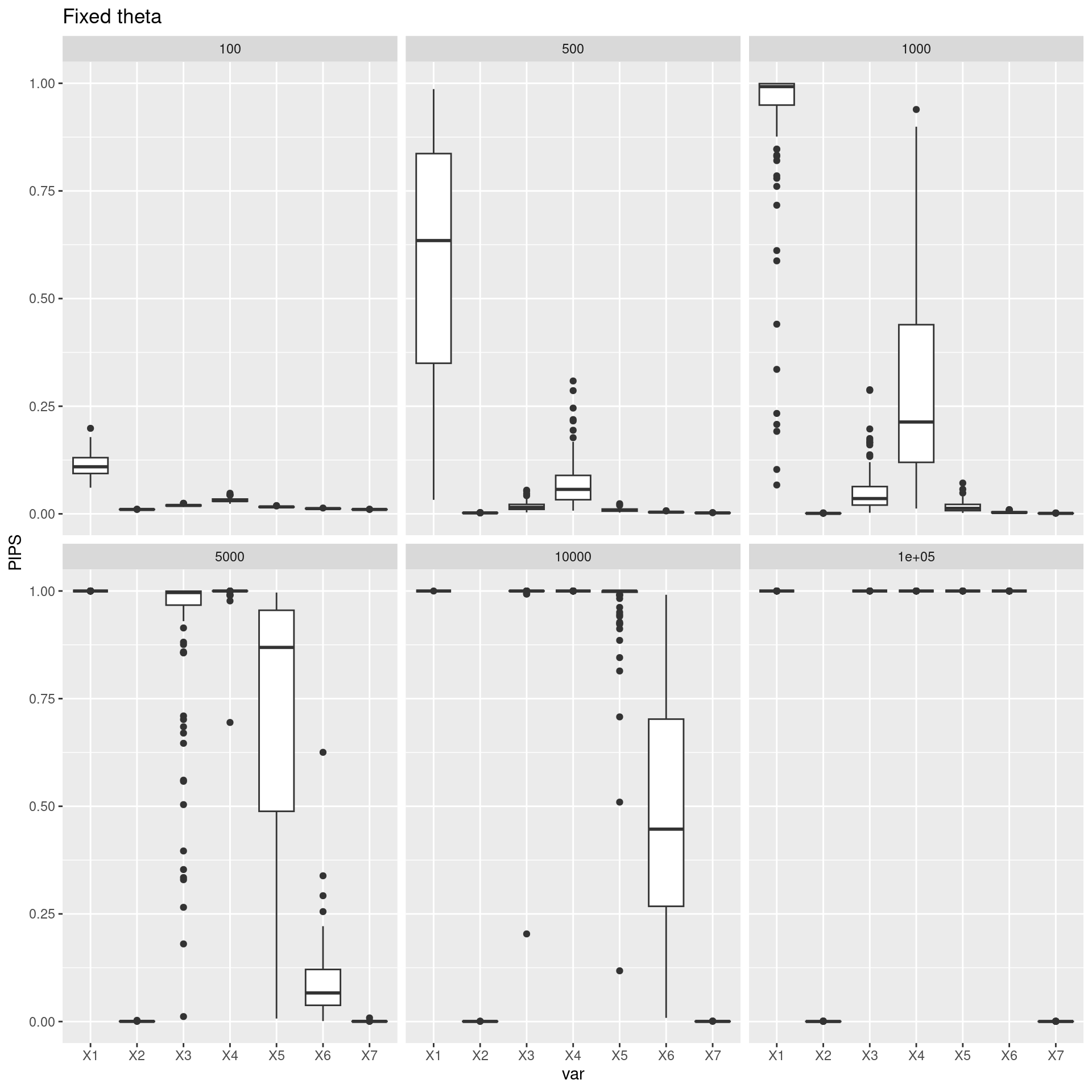}
 \caption{Posterior inclusion probabilities computed with different values of $\alpha=100;500;1000;5000;10,000;100,000$ when $\boldsymbol\theta$ is fixed to its true value.}
 \label{figalphaf}
\end{figure}

\begin{figure}
 \centering
 \includegraphics[scale=.6]{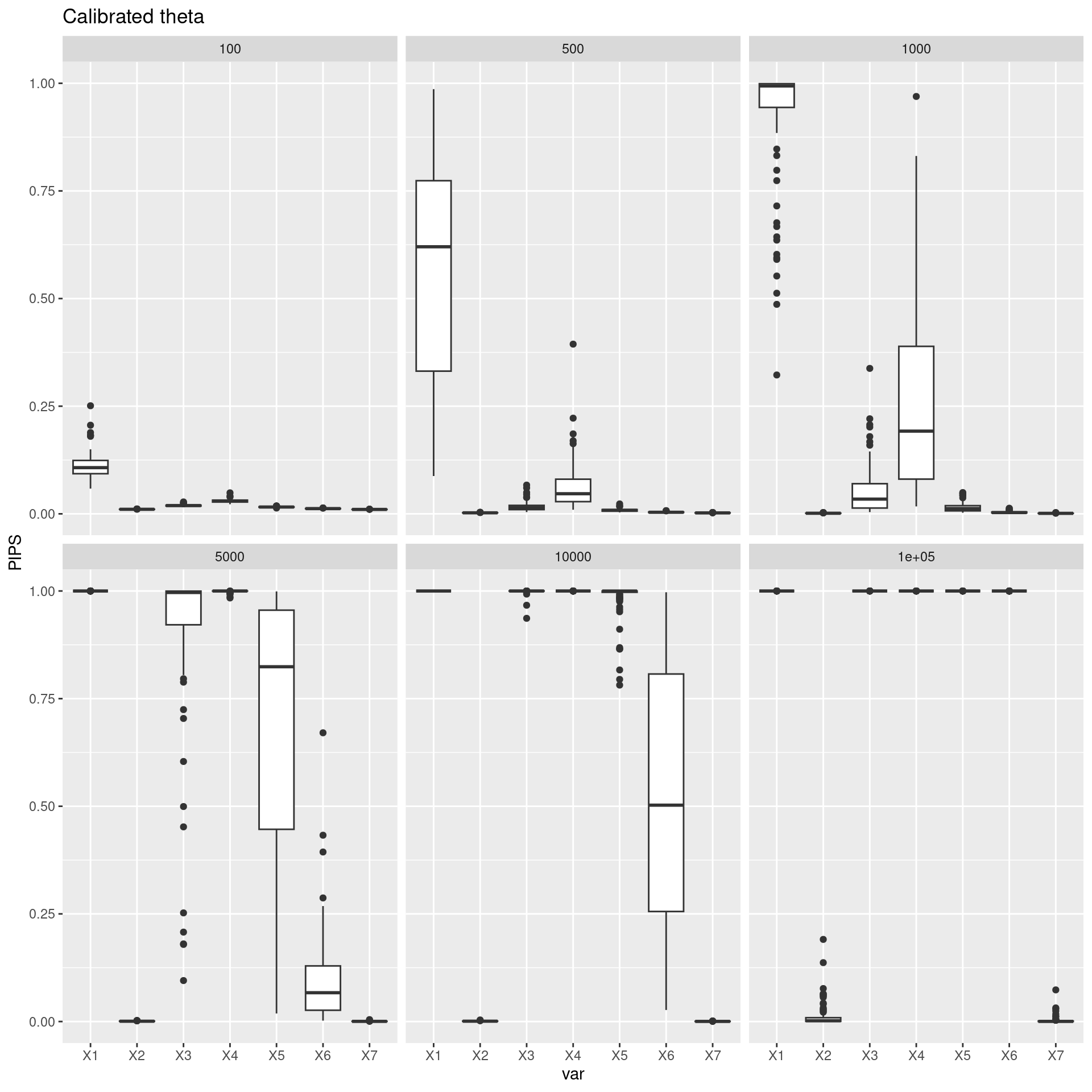}
 \caption{Posterior inclusion probabilities computed with different values of $\alpha=100;500;1000;5000;10,000;100,000$ when $\boldsymbol\theta$ is calibrated.}
 \label{figalphac}
\end{figure}

We replicate this setting 100 times with $\boldsymbol\theta$ being fixed to its true value or being calibrated.
The results for different values of $\alpha$ are displayed in Figures \ref{figalphaf} and \ref{figalphac}.
We notice that from values of $\alpha$ larger than $1000$ we start to detect all the active variables. 
The choice $\alpha=100,000$ works well but could lead to more numerical instability. Along with the heuristic argument provided in Section 3.4.4.of the paper, this is another reason why we decided to recommend $\alpha=5000$.

The R code to reproduce this example is provided in \texttt{SI4.html}.

\section{Comparison between RDVS and PIPS}\label{sec:simulated}

We consider a scenario where we have $p=8$ input variables. The $n=50$ experimental units will be chosen as a Latin-Hypercube-Design (LHD) and denoted as $\n x_i=(x_{1i},x_{2i,} \dots, x_{pi})^\top \in [0,1]^p$, for $i=1,\ldots,n$.  

The field observations will  be generated from the model
$$y_i = \CM(\n x_i,\btheta) + \delta(\n x_i) + \varepsilon_i$$
where
$$\CM(\n x_i,\btheta) = \sum_{\ell=1}^4 \frac{|4 x_{\ell i} -2| + \theta_\ell}{1+\theta_\ell}\ . $$

The four parameters $(\theta_j$, $j=1,\ldots,k=4)^\top$ are
fixed at $(0.3,0.4,0.5,0.6)^\top$. In the statistical analyses that will follow, these parameters will be considered either known and fixed 
at their true value, or as unknown, and in this case they will be calibrated.

The expression for the discrepancy function is  
\begin{equation}\label{bias}
\delta(\n x_i) = \sin(2\pi x_{1i}\cdot  x_{5i}) + x_{2i}^3 + (1-x_{6i})^3\ ,
\end{equation}
which crucially depends only on $4$ of the $8$ original input variables,  namely on $x_1$, $x_2$, $x_5$ and $x_6$. Finally, the measurement error was simulated  as $ \varepsilon_i\sim\mathcal{N}(0,\sigma_0^2=0.05^2)$. 

Notice that in this scenario we have variables which are both in the model and in the discrepancy function ($x_1$ and $x_2$), variables which only appear in the model ($x_3$ and $x_4$), variables which are only part of the discrepancy function ($x_5$ and $x_6$) and, finally, variables which do not appear neither in the model not in the discrepancy function ($x_7$ and $x_8$).  

We used the model above to generate $100$ datasets. We then applied the  proposed PIPS approach
as well as the RDVS method to identify the variables which are active in 
the discrepancy function. As mentioned, we consider $\btheta$ to be fixed at its true value or 
unknown and hence calibrated. Notice that the discrepancy function is treated as unknown (i.e., the formula \eqref{bias} used to generate the data is not used in the analysis) and is modeled
as a GaSP as in Equation (2). The measurement error variance $\sigma_0^2$ is also treated as unknown. 

For the two methods, the MCMC algorithm consists in a combination of a Metropolis within Gibbs (MwG) algorithm followed
by a Metropolis-Hastings (MH) algorithm.
First, $5,000$ iterations of an MwG algorithm are run. Then, the covariance matrix of the posterior distribution
of the parameters is estimated from this sample and is used in order to set 
the covariance of the proposal distribution of an MH algorithm. This MH algorithm is run 
for $10,000$ iterations. The proposal distributions are Gaussian random walks in 1 dimension for the MwG and in all
dimensions for the MH algorithm. The parameters are transformed to lie in a non-constrained real interval:
the variance parameters are then taken in the logarithm scale and the logit function is used to map the
the model parameters to calibrate and the $\n\rho$ parameters (all lying in the unit interval) to  
the real interval.

For the PIPS method, this algorithm is run only once, while it is repeated $T=100$ times for the RDVS method, in order to learn the distribution of the posterior median for an inert input variable. 

We fix the parameter $a$ in Equation (6) at $a=1.9$. 
Regarding the prior distributions, both methods require a prior for the variances, which we specify as inverse gamma: $\sigma^2\sim \mathcal{IG}(3,1)$ and $\sigma^2_0\sim \mathcal{IG}(4,0.02)$. This implies that, \emph{a priori}, we expect the variance of the discrepancy term to be larger than the measurement error variance. 
When the parameter $\n \theta$ is treated as unknown, a uniform distribution in the interval $[0,1]$ is set independently for each of its dimensions.
The parameters in $\n\rho$ have all the slab part of the prior distribution since the MCMC algorithm is run
only under the model $\mathcal{M}_{\boldsymbol 1}$. For the RDVS method, we also used this prior distribution. As in the previous sections, we 
chose to compute the posterior inclusion probabilities with $\alpha=5000$.


For the RDVS method, the posterior median of the range parameter corresponding to each input variable is compared to a quantile of the posterior median of the range parameter of the fictitious variable. In particular, the three percentiles used are $5\%$, $10\%$ and $15\%$. Notice that \citet{linketal2006} argue for using a relatively large percentile in order to avoid  missing an active variable; accordingly, they recommend the 10th percentile. 
 
 For the PIPS method, the posterior probability of activeness is compared to a threshold. The natural threshold is $0.5$ but, if one wants to be more
 conservative, one can choose a smaller one. In particular, we considered $0.1$, $0.5$ and $0.9$ and found the results to be not particularly sensitive to this choice in this example. 
 
 
 Tables \ref{tab:withoutcal} and \ref{tab:withcal} display the proportions of simulations where the variables are identified as active according to the two methods in competition under different settings. Table \ref{tab:withoutcal} corresponds to the case where the parameter $\btheta$ in the computer model is fixed at its true value, and Table \ref{tab:withcal} corresponds to the case where this parameter is simultaneously calibrated.

 \begin{table}
\centering
\begin{tabular}{llaarraarrr}
 \hline
 && $x_1$ & $x_2$ &$x_3$ & $x_4$ & $x_5$ & $x_6$& $x_7$ & $x_8$ \\ 
  \hline
\multirow{3}{*}{RDVS}&q5\% & 1.00 & 1.00 & 0.10 & 0.03 & 1.00 & 1.00 & 0.05 & 0.00 \\ 
 & q10\% & 1.00 & 1.00 & 0.15 & 0.05 & 1.00 & 1.00 & 0.07 & 0.00 \\ 
  &q15\% & 1.00 & 1.00 & 0.18 & 0.05 & 1.00 & 1.00 & 0.07 & 0.00 \\ 
  \hline
  \multirow{3}{*}{PIPS}&th0.1 & 1.00 & 1.00 & 0.00 & 0.00 & 1.00 & 1.00 & 0.00 & 0.00 \\ 
  &th0.5 & 1.00 & 1.00 & 0.00 & 0.00 & 1.00 & 1.00 & 0.00 & 0.00 \\ 
  &th0.9 & 1.00 & 1.00 & 0.00 & 0.00 & 1.00 & 1.00 & 0.00 & 0.00 \\ 
  \hline
\end{tabular}
\caption{Proportion of detection for a variable to be active when using RDVS and PIPS methods when the parameters $\btheta$ are fixed to their true values.
For RDVS, the  posterior median is compared with quantiles $q=5\%,10\%,15\%$ of the posterior median of the fictitious variable. For PIPS, the tested thresholds are fixed to $th=0.1,0.5,0.9$. Shadowed columns represent the truly active variables in the discrepancy function.} 
\label{tab:withoutcal}
 \end{table}

\begin{table}
\centering
\begin{tabular}{llaarraarrr}
 \hline
 && $x_1$ & $x_2$ &$x_3$ & $x_4$ & $x_5$ & $x_6$& $x_7$ & $x_8$ \\ 
  \hline
\multirow{3}{*}{RDVS}&q5\% & 1.00 & 1.00 & 0.03 & 0.03 & 1.00 & 1.00 & 0.03 & 0.00 \\ 
  &q10\% & 1.00 & 1.00 & 0.07 & 0.05 & 1.00 & 1.00 & 0.03 & 0.00 \\ 
  &q15\% & 1.00 & 1.00 & 0.12 & 0.05 & 1.00 & 1.00 & 0.03 & 0.00 \\
  \hline
  \multirow{3}{*}{PIPS}&th0.1 & 1.00 & 1.00 & 0.00 & 0.00 & 1.00 & 1.00 & 0.00 & 0.00 \\ 
  &th0.5 & 1.00 & 1.00 & 0.00 & 0.00 & 1.00 & 1.00 & 0.00 & 0.00 \\ 
  &th0.9 & 1.00 & 1.00 & 0.00 & 0.00 & 1.00 & 1.00 & 0.00 & 0.00 \\ 
  \hline
\end{tabular}
\caption{Proportion of detection for a variable to be active when using RDVS and PIPS methods when the parameters $\btheta$ are calibrated.
For RDVS, the  posterior median is compared with quantiles 
 $q=5\%,10\%,15\%$ of the posterior median of the fictitious variable. For PIPS, the tested thresholds are fixed to $th=0.1,0.5,0.9$. Shadowed columns represent the truly active variables in the discrepancy function.} 
\label{tab:withcal}
 \end{table}
Both methods show convincing results regardless of whether the calibration was performed or not. The RDVS method may suffer from some false detection  of active variables which increases as the quantile increases, as it may be expected.  Hence, the PIPS method leads to results at least as good as the RDVS method but  only requires one MCMC sample instead of one hundred. Furthermore, the PIPS method is stable and does not appear to be too sensitive to the choice of the threshold.

\textsl{\bibliographystyle{chicago}
\bibliography{biblio}}